\documentclass[useAMS,usenatbib,twocolumn]{aastex63}
\usepackage{times}
\usepackage{epsfig}
\usepackage{amssymb}
\usepackage{amsmath}

\begin{document}

\title[Observational limits on the early-time dust mass in SN~1987A]
{Observational limits on the early-time dust mass in SN~1987A}

\correspondingauthor{Roger Wesson}
\email{rw@nebulousresearch.org}

\author[0000-0002-4000-4394]{Roger Wesson}
\affiliation{Department of Physics and Astronomy, University College London, Gower Street, London WC1E 6BT, United Kingdom}
\affiliation{The authors contributed equally to this work.}
\author[0000-0001-7188-6142]{Antonia Bevan}
\affiliation{Department of Physics and Astronomy, University College London, Gower Street, London WC1E 6BT, United Kingdom}
\affiliation{The authors contributed equally to this work.}

\date{Received:}


\begin{abstract}

In recent years, dust masses of a few tenths of a solar mass have been found in the expanding ejecta of a number of core-collapse supernovae. How dust forms in such quantities remains poorly understood; theories of dust formation predict lower total masses and much faster formation rates than observations imply. One suggestion to reconcile observations and theory was made by \citet{dwek2019}, who proposed that the dust forms very rapidly, and because of its optical depth, is not initially observationally detectable, only being gradually revealed as the ejecta expand. Observational dust masses at early times would then only be lower limits. Using a large grid of radiative transfer models covering dust masses from 10$^{-4}$ to 1\,M$_\odot$ to calculate both the spectral energy distribution and the emission line profiles from clumpy dust shells, we show that this cannot be the case. Some clump distributions allow dust masses of $\sim$0.01\,M$_\odot$ to be concealed in clumps and still predict an SED consistent with the observations. However, these geometries predict emission line profiles which are inconsistent with the observations. Similarly, clump geometries which reproduce the observed emission line profiles with dust masses $>$0.01\,M$_\odot$ do not reproduce the SED. However, models with $\sim$10$^{-3}$\,M$_\odot$ of amorphous carbon can reproduce both the SED and the emission line profiles. We conclude that no large masses of dust can be hidden from view in the ejecta of SN\,1987A at early epochs, and that the majority of dust must thus have formed at epochs $>$1000 days.

\end{abstract}



\section{Introduction}

Since the discovery that galaxies at high redshifts already contain large quantities of dust (\citealt{bertoldi2003,laporte2017,hashimoto2018}) the origins of that dust have been the subject of vigorous research. While theory predicted that dust should form in significant quantities in the expanding and cooling ejecta of core-collapse supernovae (CCSNe; see e.g. \citealt{todini2001}), a number of observational studies of CCSNe at ages of a few hundred days found quantities of dust too small to resolve the early universe dust budget problem (e.g. \citealt{meikle2011}). However, the 2011 discovery of several tenths of a solar mass of dust in the remnant of SN~1987A, 23 yr after its explosion (\citealt{matsuura2011}) confirmed that supernovae could indeed produce dust in the necessary quantities.

Determining when this dust forms is then key to understanding why supernova remnants seem to contain low dust masses a few hundred days after explosion, but high dust masses years, decades or centuries later, as also observed in Cas A (\citealt{barlow2010,arendt2014,bevan2017,delooze2017}), the Crab nebula (\citealt{gomez2012,owen2015,delooze2019}), SN~1980K and SN~1993J (\citealt{bevan2017}), G11.2-0.3, G21.5-0:9
and G29.7-0:3 (\citealt{chawner2019}; \citealt{priestley2020}), and G54.1+0.3 (\citealt{temim2017}, \citealt{rho2018}, \citealt{priestley2020}). Recent theoretical work has found that dust formation in CCSN ejecta should be complete within 1000 days of the explosion (\citealt{sarangi2015}, \citealt{sluder2018}). However, several observational studies have presented evidence that dust formation must continue for years (\citealt{gall2014}, \citealt{wesson2015}, \citealt{bevan2016}). Additional evidence for slow dust formation comes from studies of presolar dust grains in meteorites: \citet{liu2018} found that $^{49}$Ti and $^{28}$Si isotope ratios were consistent with dust forming after a significant fraction of $^{49}$V (half-life 330 days) had decayed into $^{49}$Ti, while \citet{ott2019} found that barium isotope ratios were consistent with silicon carbide dust grains having formed when a substantial fraction of $^{137}$Cs (half-life 30 yr) had decayed into $^{137}$Ba. Although the extraction of isotope ratios from presolar grains and the nucleosynthetic model adopted are both sources of significant uncertainty, these results also indicate that the bulk of the dust formation occurs years to decades after the supernova outburst.

An alternative scenario has been presented by \citet{dwek2015} and \citet{dwek2019}, who have proposed that the formation of large quantities of silicate dust occurs within a few hundred days, but within such optically thick clumps that it radiates only at extremely long wavelengths and thus escapes detection in observations at shorter wavelengths. Many models of supernova ejecta dust find that carbon dust is necessary to explain largely featureless SEDs, in tension with the expectation that the progenitors of CCSNe should generally be oxygen-rich. A high mass of silicate dust in dense clumps could also resolve that tension, since the high optical depths involved would obscure the silicate emission features, giving the required featureless SED. If a high dust mass could be concealed at early times, this would also conform with the timescales for dust formation found in theoretical studies.

Clumping in supernova ejecta is expected to arise from Rayleigh-Taylor instabilities at the onset of the explosion (\citealt{arnett1989}), and is well supported observationally (\citealt{lucy1991}, \citealt{biermann1990}, \citealt{dessart2018}). \citet{ercolano2007a}, \citet{wesson2015} and \citet{bevan2016} have considered clumpy models of the dust in the ejecta of SN~1987A, using both spectral energy distribution and emission line profile modeling; none of these studies identified any parameters which were able to conceal large quantities of dust in the ejecta at early times. In this work, we expand the parameter space investigated in those papers, and search for dust configurations that can reproduce both the SED and the emission line profiles, to further constrain how much dust could be present before SN~1987A was 1000 days old.

\section{Radiative transfer models}

Dust masses in supernova remnants can be estimated by fitting the optical-submillimeter spectral energy distribution (SED). This approach has been widely used over many years (e.g. \citealt{kotak2005}, \citealt{sugerman2006}, \citealt{ercolano2007a}, \citealt{wesson2010a}, \citealt{meikle2011}, \citealt{wesson2015}). If the observational data does not extend to far-infrared wavelengths where cold dust emission peaks, it is possible for cold dust to escape detection. However, such dust will still be detectable through its effect on optical emission line profiles. Emission from the far side of an expanding dusty shell will experience extinction which emission from the near side is not subjected to, and thus, the redward side of emission line profiles is preferentially absorbed. In recent years, this technique has been used to estimate dust masses, which have been found to be in good agreement with those independently derived from SED fitting in several objects (\citealt{bevan2016}, \citealt{bevan2017}, \citealt{bevan2018}, \citealt{bevan2019}). By applying both SED fitting and emission line fitting, one can place constraints on dust properties that are not available from either technique alone.

In this paper, we create models applicable to the remnant of SN~1987A at an epoch of $\sim$800 days. We chose this epoch because of the availability of observations constraining both the optical-far IR SED and the optical emission line profiles. The SED at this epoch consists of ground-based optical spectrophotometry from  \citet{phillips1990}, Kuiper Airborne Observatory (KAO) near-infrared and mid-infrared spectrophotometry from \citet{wooden1993} and \citet{dwek1992} respectively, and far-infrared KAO photometry from \citet{harvey1989}, consisting of a detection at 50$\mu$m and an upper limit at 100$\mu$m. The optical spectrophotometric observations at 806 days obtained by \citet{phillips1990} at the Cerro Tololo Inter-American Observatory (CTIO) 1.5m telescope show strong, blueshifted [O~{\sc i}]~6300,6363~\AA\ emission, allowing the application of emission-line fitting to estimate the dust mass independently of the SED.

\subsection{The SED model}
\label{mocassin}

Estimates of the dust mass in SN\,1987A using three-dimensional radiative transfer models have previously been presented by \citet{ercolano2007a} and \citet{wesson2015}. These authors found dust masses at 775 days, assuming amorphous carbon dust grains, of 4.2$\times$10$^{-4}$ and 2$\times$10$^{-3}$\,M$_\odot$ respectively.

We calculated spectral energy distributions for model dust shells using the three-dimensional radiative transfer code {\sc mocassin} v2.02.73 (\citealt{ercolano2003}, \citealt{ercolano2005}). We created a grid of models, based on those of \citet{ercolano2007a} and \citet{wesson2015} but covering an expanded parameter space. Compared to the range investigated by \citet{ercolano2007a} and \citet{wesson2015}, we model a greater range of dust masses (from 10$^{-4}$ to 1\,M$_\odot$); a much wider range of clump geometries, described by filling factors from 0.005-0.5, clump radii of R$_{\rm out}$/15 -- R$_{\rm out}$/45 and radial number densities varying with r$^{0}$ to r$^{-4}$; a range of grain radii from 0.005 to 1$\mu$m, where the earlier works considered only MRN grain size distributions; and three dust species: amorphous carbon and astronomical silicates as modeled in the earlier works, and additionally MgSiO3 which was proposed by \citet{dwek2015} and \citet{dwek2019} to have formed in significant quantities by the epoch of interest.

The method of selection of the parameter ranges we explored is described below. For all our models, we adopted a distance to the LMC of 49.97 kiloparsecs (\citealt{pietrzynski2013}).

\begin{table}
\caption{The fixed parameters for the SED models.}
\begin{tabular}{ll}
Parameter & Values \\
\hline
Luminosity & 1.6$\times$10$^{5}$~L$_\odot$ \\
Temperature & 7000~K \\
R$_{\rm{in}}$ & 6.3$\times$10$^{15}$~cm \\
R$_{\rm{out}}$/R$_{\rm{in}}$ & 5 \\
\end{tabular}
\label{SED_fixed_params}
\end{table}

\begin{table}
\caption{The variable parameter space covered by our grid of radiative transfer models. *Optical constants from \citet{zubko1996}, \citet{draine1984} and \citet{jaeger2003} for amorphous carbon, silicates and MgSiO$_3$ respectively.}
\begin{tabular}{ll}
Parameter & Values \\
\hline
Clump filling factor ($f$) & 0.005, 0.01, 0.025, 0.05, 0.1, 0.2, 0.5 \\
Clump radius (R$_c$) & R$_{\rm{out}}$/15, R$_{\rm{out}}$/30, R$_{\rm{out}}$/45 \\
Clump number density $\propto$r$^-$ & 0, 2, 4 \\
Dust composition & amorphous carbon,\\
                 & silicates, MgSiO$_3^*$ \\
Grain radius ($a$; $\mu$m) & 0.005, 0.01, 0.0316, 0.1, 0.316, 1.0 \\
Dust mass (M$_{\rm dust}$; M$_\odot$) & 10$^{-4}$, 10$^{-3}$, 10$^{-2}$, 10$^{-1}$, 1.0 \\
\hline
\end{tabular}
\label{parameterspace}
\end{table}

\begin{table}
\caption{The fixed parameters for the emission line profile models.}
\begin{tabular}{ll}

Parameter & Values \\
\hline
Maximum velocity & 4000~km~s$^{-1}$ \\
R$_{\rm{out}}$ & 2.8$\times$10$^{16}$~cm \\
R$_{\rm{out}}$/R$_{\rm{in}}$ & 16.7 \\\relax
[O~{\sc i}] doublet ratio & 2.3 \\
emissivity & $\propto r^{-4.8}$
\end{tabular}
\label{line_fixed_params}
\end{table}

\subsubsection{Model assumptions}

In the grid of models that we constructed, we held certain parameters constant. These are summarised in Table~\ref{SED_fixed_params}. The temperature and luminosity of the source are assumed to be 7000~K and 1.6$\times$10$^{5}$~L$_\odot$ respectively. The former is well determined from the optical data, and the observed bolometric luminosity (\citealt{suntzeff1991}) constrains the latter. In the case of a smooth dust distribution, there would be a degeneracy between the luminosity and the dust mass which would make the adoption of a single luminosity potentially problematic. However, for the clumpy dust geometries that we are investigating, the amount of optical emission depends on the clump filling factor rather than the dust mass. The majority of our models reproduce the average flux in the 0.3-1.0$\mu$m region to within 3$\sigma$ of the observed value, indicating that the luminosity we have adopted is a reasonable choice.

Following the models of \citet{ercolano2007a} and \citet{wesson2015}, the inner and outer radii are fixed at values of 6.3$\times$10$^{15}$cm and 31.5$\times$10$^{15}$cm respectively, corresponding to velocities of 940 and 4700 kms$^{-1}$. This is consistent with the values adopted by \citet{bevan2016} for their line profile models of H$\alpha$, although they found that a much smaller inner radius was necessary to fit the profile of the [O~{\sc i}] 6300,6363 doublet. This indicates that [O~{\sc i}] is concentrated toward the center of the expanding remnant, while H$\alpha$ emission is more diffuse. Because [O~{\sc i}] emission therefore has to cross a larger column depth of dust before escaping the remnant, \citet{bevan2016} found that the dust masses required to fit its profile were a factor of $\sim$4 higher than the dust masses necessary to fit the H$\alpha$ line profile.

The SED is relatively insensitive to the inner radius, as the luminosity of the heating source scales with the dust density in the interclump medium. Changing the inner radius does not result in the dust being exposed to a significantly different radiation field. To facilitate comparison with the models presented in \citet{wesson2015}, we retain the inner and outer radii used in that work. While this means that the SED fitting and line profile fitting models are not using identical dust distributions, it allows the line profile models to maximize the dust mass, in line with our goal of determining whether high dust masses as envisaged by \citet{dwek2015} and \citet{dwek2019} are feasible.

\begin{figure*}
\includegraphics[width=\textwidth]{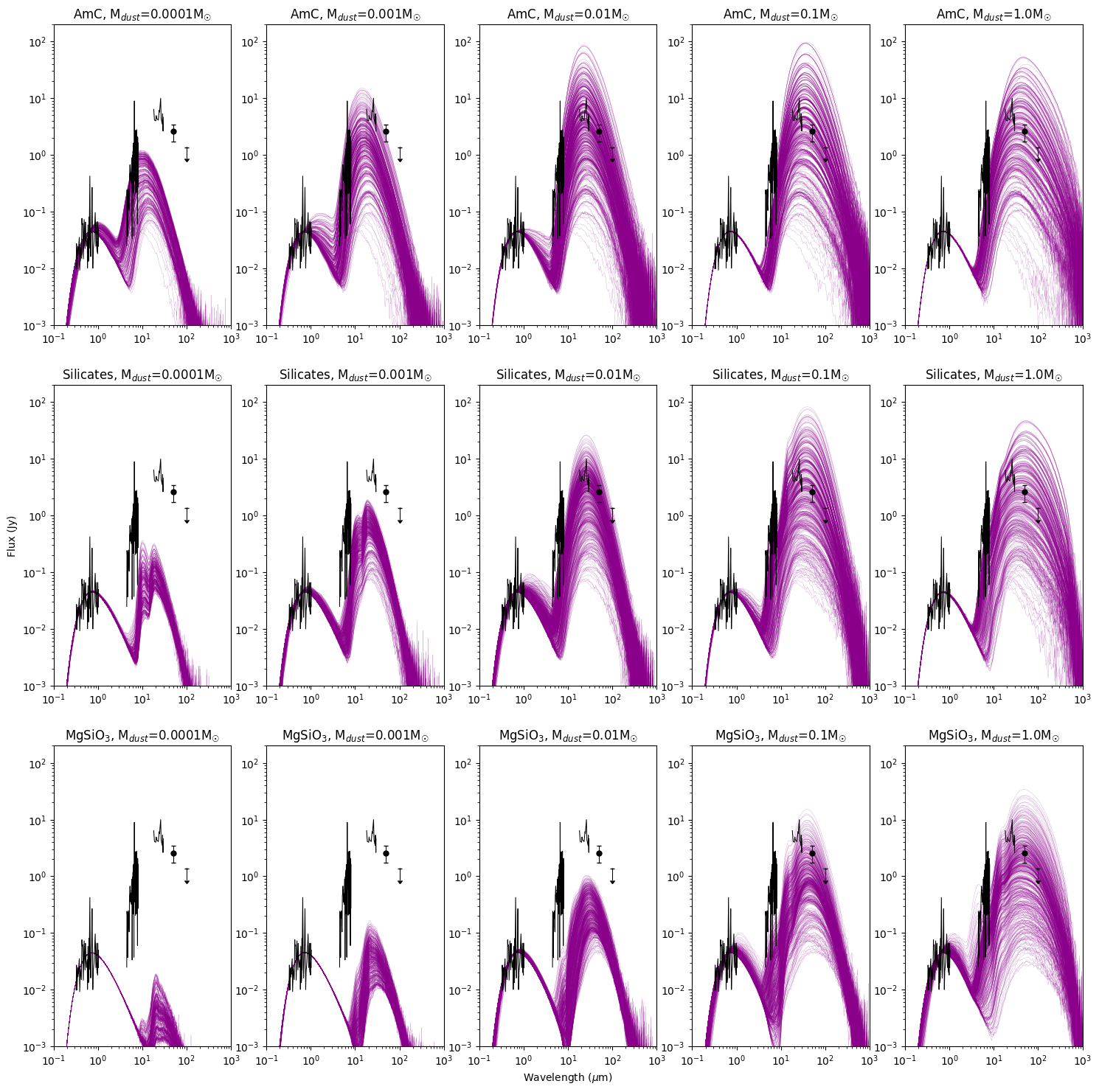}
\caption{Predicted SEDs, normalized to the observed flux at 0.7$\mu$m, for all combinations of model parameters for (top to bottom) amorphous carbon, silicates, MgSiO$_3$, and (left to right) 10$^{-4}$, 10$^{-3}$, 10$^{-2}$, 10$^{-1}$ and 1.0\,M$_\odot$ of dust. This illustrates the general dependence of the SED on species and dust mass; amorphous carbon dust can approximately fit the observations with a lower mass of dust than any silicate species, and even with a dust mass of 1\,M$_\odot$, most silicate species models do not well reproduce the near-infrared part of the SED. Optical-infrared spectrophotometry is from \citet{wooden1993} on day 775, and far-infrared photometry is from \citet{harvey1989} on day 791-795.}
\label{allseds}
\end{figure*}

\begin{figure*}
\includegraphics[width=\textwidth]{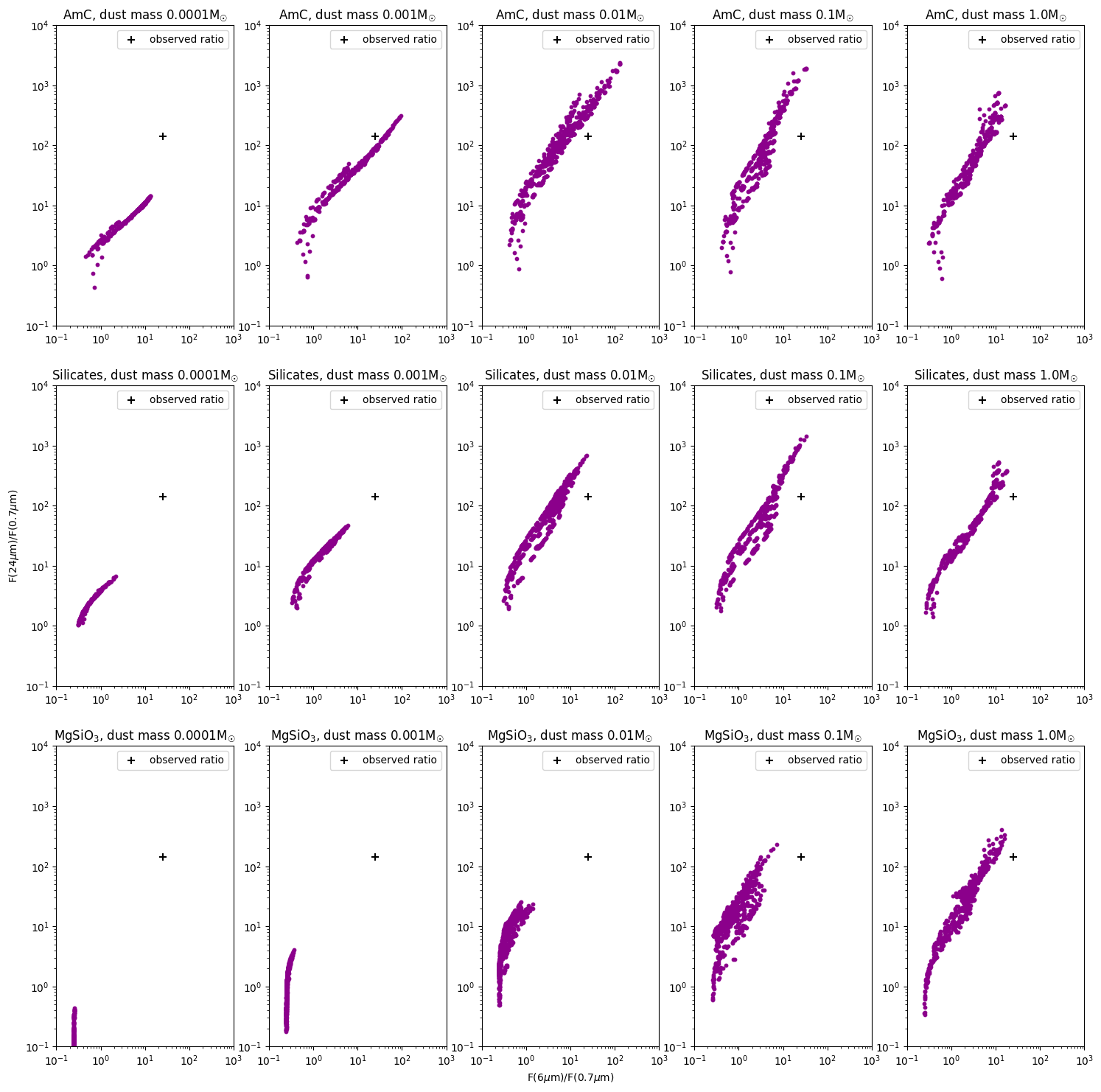}
\caption{The ratio of 24$\mu$m/0.7$\mu$m against 6.0$\mu$m/0.7$\mu$m fluxes predicted for all combinations of (top to bottom) amorphous carbon, silicates, MgSiO$_3$, and (left to right) 10$^{-4}$, 10$^{-3}$, 10$^{-2}$, 10$^{-1}$ and 1.0\,M$_\odot$ of dust, compared to the observed value at day 775.}
\label{ratiopanel1}
\end{figure*}

\begin{figure*}
\includegraphics[width=\textwidth]{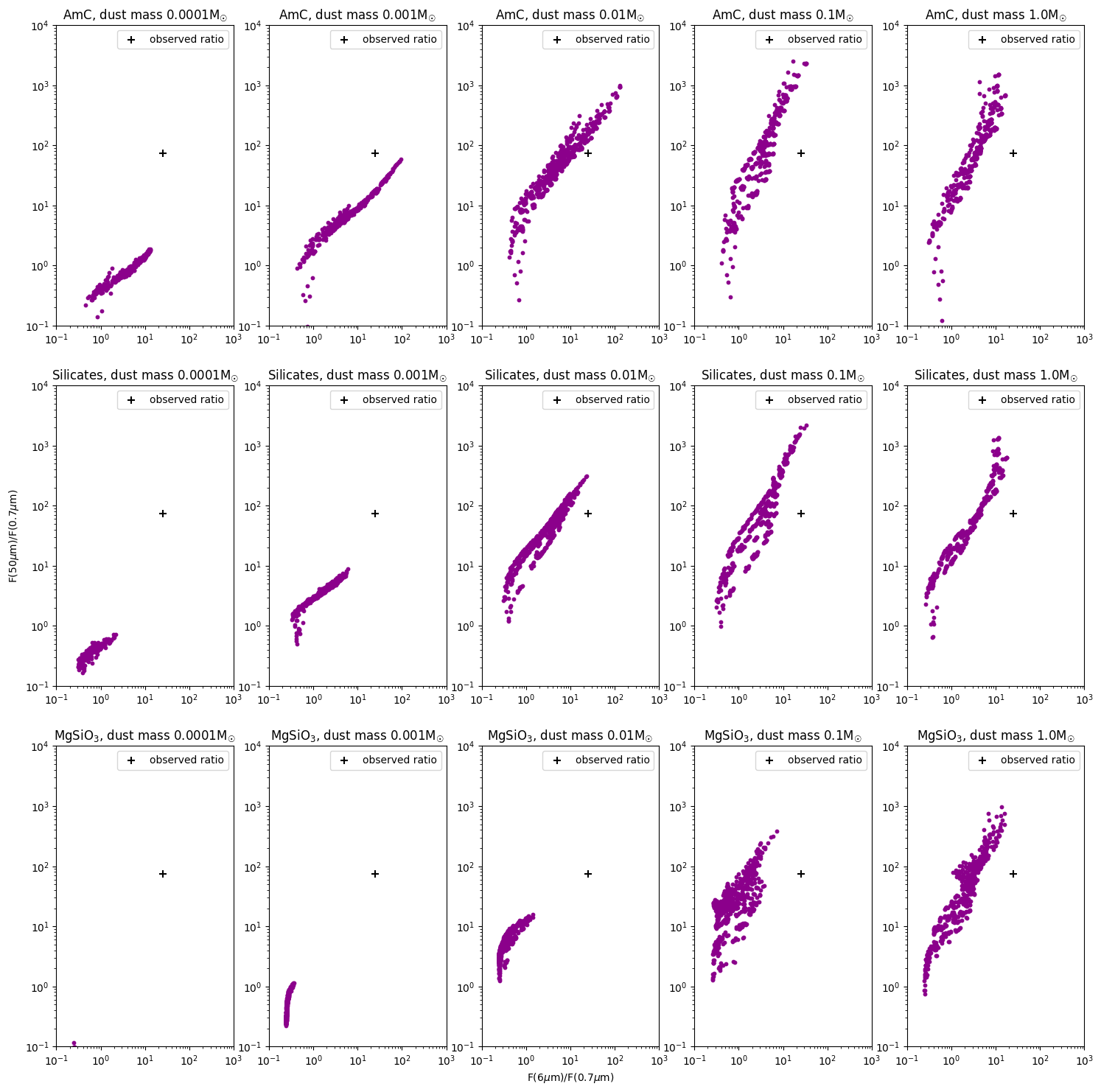}
\caption{The ratio of 50$\mu$m/0.7$\mu$m against 6.0$\mu$m/0.7$\mu$m fluxes predicted for all combinations of (top to bottom) amorphous carbon, silicates, MgSiO$_3$, and (left to right) 10$^{-4}$, 10$^{-3}$, 10$^{-2}$, 10$^{-1}$ and 1.0\,M$_\odot$ of dust, compared to the observed value at day 775.}
\label{ratiopanel2}
\end{figure*}

\begin{figure*}
\includegraphics[width=\textwidth]{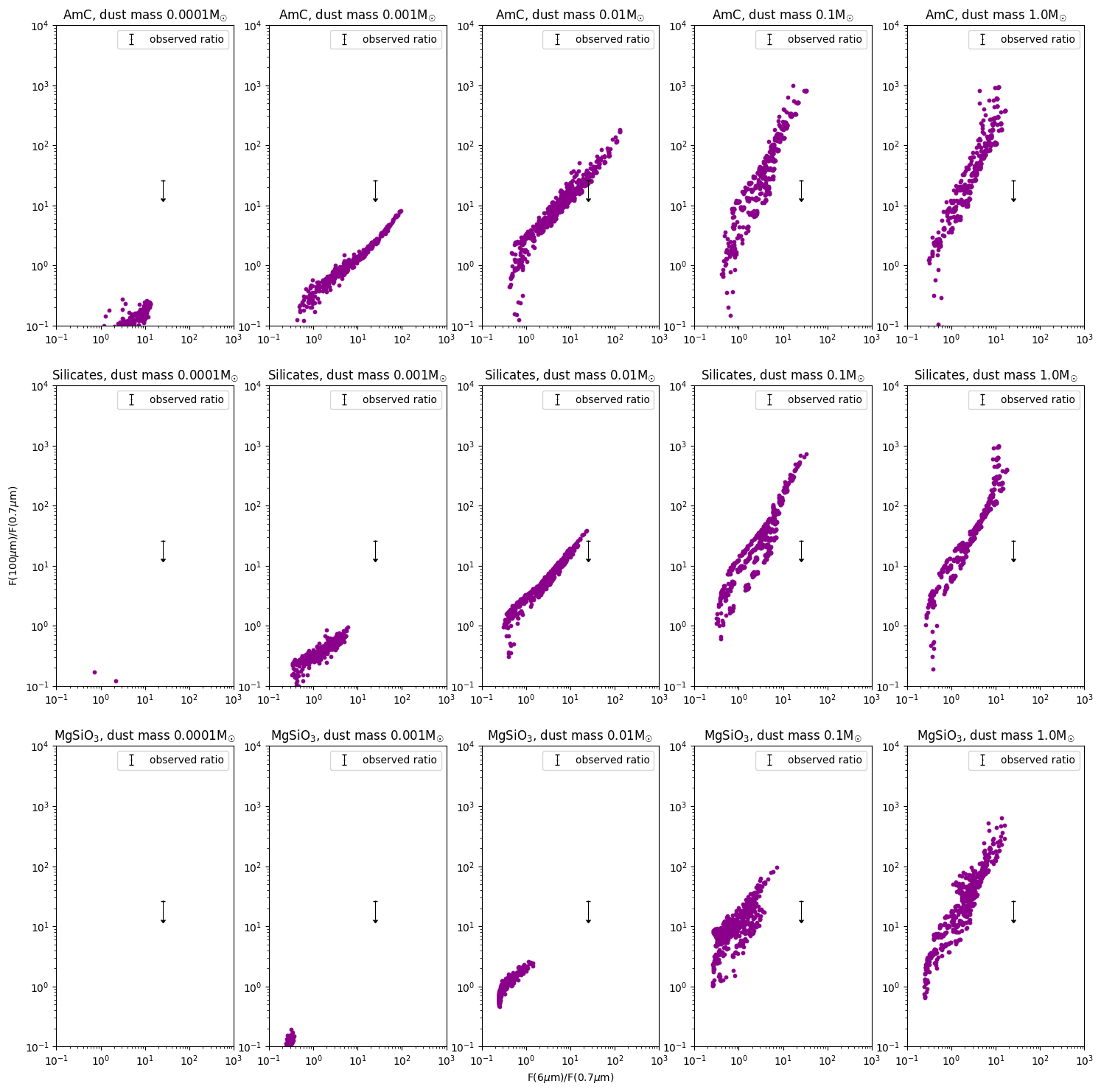}
\caption{The ratio of 100$\mu$m/0.7$\mu$m against 6.0$\mu$m/0.7$\mu$m fluxes predicted for all combinations of (top to bottom) amorphous carbon, silicates, MgSiO$_3$, and (left to right) 10$^{-4}$, 10$^{-3}$, 10$^{-2}$, 10$^{-1}$ and 1.0\,M$_\odot$ of dust, compared to the observed value at day 775.}
\label{ratiopanel3}
\end{figure*}

\subsection{The emission line profile model}

The optical spectrum of SN~1987A at 806 days showed strong H$\alpha$ and [O~{\sc i}]~6300,6363~\AA\ emission with the [O~{\sc i}] doublet exhibiting a more extreme blueshifted asymmetry than the H$\alpha$ line. We focused modeling on the [O~{\sc i}]~6300,6363~\AA\ doublet since the more extreme asymmetry is a weaker constraint on the maximum dust mass possible in the ejecta at this time.

We created a grid of models based on the work of \citet{bevan2016} but covering an expanded parameter space specifically exploring a wider range of clump filling factors, clump sizes and clump number densities to investigate their effects on the modeled doublet. A wider dust mass range spanning dust masses from $10^{-4}$ to 1.0\,M$_{\odot}$ was explored with the aim of determining whether large masses of dust hidden in optically thick clumps could be consistent with the observed [O~{\sc i}]~6300,6363~\AA\ doublet at  806 days. The code {\sc damocles} \citep{bevan2016} was used to create the modeled line profiles. The grid of variable parameters that was explored was identical to the grid for the SED models presented in section \ref{mocassin} (see Table \ref{parameterspace}).

\subsubsection{Model assumptions}

Certain parameters were fixed in order to reduce the potential parameter space and focus on the effects of clumping and geometry on the dust mass required to reproduce the line profile. Values were adopted for these parameters based on the best-fitting models of the [O~{\sc i}]~6300,6363~\AA\ doublet presented in \citet{bevan2016}. These were the maximum velocity of the ejecta (and hence the outer radius derived as $R_{out} = v_{max}t$), the ratio of the outer radius to the inner radius, the steepness of the smooth emissivity distribution and the flux ratio of the doublet components. The fixed parameters are summarized in Table \ref{line_fixed_params}.

\subsection{The variable parameters}
\subsubsection{Clump filling factor}

Our {\sc mocassin} models adopt a configuration in which all of the dust is contained in clumps, and all of the heating comes from the interclump medium. \citet{ercolano2007a} found that such a configuration maximizes the dust mass required to fit the SED. In this configuration, when the clumps become optically thick, the optical extinction is a function only of the volume filling factor of the clumps, while the IR excess is a function only of the dust mass. Fitting an SED then becomes a matter of determining the clump filling factor necessary to fit the optical part of the SED, and separately determining the dust mass necessary to fit the IR. \citet{wesson2015} investigated four clump filling factors, $f$=0.025, 0.05, 0.1 and 0.2. We extend that range to include filling factors of $f$=0.005, 0.01 and 0.5. We do not consider the limiting case of a filling factor of 1.0, i.e. a smooth dust distribution; the observed optical emission constrains the dust mass in this case to around $\sim$10$^{-4}$\,M$_\odot$ (\citealt{wooden1993}) and so a large dust mass is already definitively ruled out in this case.

\subsubsection{Clump radius}

The origin of clumping in supernova remnants is assumed to be Rayleigh-Taylor instabilities developing soon after the shock breakout. These are predicted to have a size of 1-3\% of the ejecta radius \citep{arnett1989b}. \citet{wesson2015} investigated clump radii of R/30, R/45 and R/60, but with little exploration of the smaller values due to computational limitations. Improvements to {\sc mocassin}'s performance since then allowed us to carry out a more detailed investigation. R/60 models remain prohibitively time-consuming, but we ran a full grid of models with clump sizes of R/15, R/30 and R/45.

\subsubsection{Clump number density}

\citet{wesson2015} considered a clump distribution with a number density proportional to r$^{-2}$, with limited investigations of shallower and steeper density distributions. In this work, our grid of models also fully covers density distributions proportional to r$^{0}$ and r$^{-4}$. Even steeper density distributions have been inferred for the ejecta of core-collapse supernovae; \citet{anderson2014} adopt an exponent of -8 to represent the ejecta density outside $\sim$3,000 km\,s$^{-1}$. We are not able to consider such steep distributions due to computational constraints: the grid resolution required to properly represent a clumpy distribution increases with the exponent of the radial density distribution, and for exponents steeper than -4, the runtime becomes prohibitive.

Qualitatively considering the effect of steeper density distributions on the SED, one would expect that a steeper distribution would eventually result in too much optical radiation escaping, as the outer regions become essentially empty while all the dust is concentrated in an increasingly small central part of the ejecta. The dust mass could be unconstrained by the optical-IR SED, but it would not be possible to correctly fit the optical part of the SED. We therefore do not expect that steeper density distributions could provide a mechanism to conceal more dust.

Similarly, the radiation emitted by the gas must be exposed to a minimum covering fraction of dust in order to induce an asymmetry in the observed line profile. Were the dust to be concentrated in a very small region in the center of the ejecta, the majority of the more diffuse [O~{\sc i}]~6300,6363\AA\ doublet would be able to escape the ejecta without interaction resulting in a (nearly) symmetric line profile shape.

\subsubsection{Dust composition}

The featureless IR SED of SN~1987A has led several authors to conclude that the dust composition must be predominantly carbonaceous (\citealt{wooden1993}, \citealt{ercolano2007a}, \citealt{wesson2015}). The absence of silicate features has been noted in a number of other supernova remnants (\citealt{fox2010}, \citealt{vandyk2013}, \citealt{williams2015}). This is somewhat puzzling as the progenitors of core-collapse supernovae would generally be expected to be oxygen-rich at the time of the explosion. \citet{dwek2015} proposed that the ejecta could be silicate-rich and fit a featureless SED if the optical depth and hence dust mass is very high. We thus consider both amorphous carbon and silicate dust compositions. For amorphous carbon, we use the optical constants of \citet{zubko1996}, while for silicates, we calculate SEDs for two varieties: astronomical silicates, using the optical constants of \citet{draine1984}, and MgSiO$_3$, using data from \citet{jaeger2003}. We assume grain densities of 1.85~g\,cm$^{-3}$ for amorphous carbon, 3.6~g\,cm$^{-3}$ for astronomical silicates, and 2.7~g\,cm$^{-3}$ for MgSiO$_3$.

\subsubsection{Dust mass}

We considered dust masses between 10$^{-4}$ and 1.0~M$_\odot$ at decade intervals for all species. The lowest mass is comparable to those estimated by early studies assuming smooth distributions of dust (e.g. \citealt{wooden1993}). A mass of 0.01~M$_\odot$ is already a factor of 5--25 times higher than the values found for day 775 by \citet{ercolano2007a} and \citet{wesson2015}, while a mass of 1.0\,M$_\odot$ is comparable to the largest estimates of the dust mass at very late times (\citealt{matsuura2011,matsuura2015}, \citealt{wesson2015}), and to the mass proposed by \citet{dwek2015} to have already formed after a few hundred days. Higher dust masses still are implausible given the constraints of nucleosynthetic yields (\citealt{woosley1988}, \citealt{thielemann1990}). The range also encompasses the final dust masses predicted by recent theoretical calculations (\citealt{sarangi2015}, \citealt{sluder2018}).

Theoretical models predict that a number of different dust species should form in CCSNe ejecta (\citealt{sarangi2015}, \citealt{sluder2018}). However, to avoid investigating an unmanageably large parameter space, in this study we investigate single grain species only.

\subsubsection{Grain size distribution}

\citet{ercolano2007a} and \citet{wesson2015} assumed an MRN grain radius distribution (\citealt{mathis1977}) for the dust in the ejecta of SN~1987A. This power-law distribution has its origins in collision and fragmentation over long periods of time, as experienced by dust in the interstellar medium but not necessarily by dust in unshocked supernova remnants. More recent work has consistently found evidence of large grains forming in SN ejecta (\citealt{gall2014}, \citealt{owen2015}, \citealt{wesson2015}, \citealt{bevan2016}). Emission line profile models are sensitive only to albedo and dust optical depth and therefore can only be used to determine two out of a minimum grain radius, a maximum grain radius and a power-law exponent. Rather than fixing one of these, we instead choose to investigate models using a single grain size, and we investigated radii ranging from 0.005 to 1.0$\mu$m.

In total, our {\sc mocassin} and {\sc damocles} grids consisted of 5670 individual models each. The full parameter space is given in Tables~\ref{SED_fixed_params} and \ref{parameterspace}.

\subsection{Comparison of SED model fluxes to observations}
\label{sedcomparison}

To compare our grid of models to the observed fluxes, we normalized all predicted SEDs to match the observed 0.7$\mu$m flux at day 775 from \citet{wooden1993}. To resolve the actual differences in predicted fluxes compared to the observations would require adjustment of the luminosity, and so even an apparently good fit to the data may not in practice be possible. Models that do not fit the data when normalized in this way, however, can be definitely ruled out. This simplification may therefore fail to exclude models that do not fit the data, but will not fail to include models that do fit the data. In Figure~\ref{allseds}, we show the normalized SEDs for the entire set of models.

We also constructed color-color diagrams to compare the model SEDs to the observations, using the average fluxes in the three regions of the day 775 \citet{wooden1993} spectrophotometry (0.32-1.03$\mu$m, 4.47-8.1$\mu$m and 18.15-29.63$\mu$m), and the \textit{Kuiper Airborne Observatory} (KAO) 50 and 100$\mu$m fluxes. We refer to the three \citet{wooden1993} spectral ranges by their average wavelengths, 0.7$\mu$m, 6$\mu$m and 24$\mu$m. Figures~\ref{ratiopanel1}, \ref{ratiopanel2} and \ref{ratiopanel3} show respectively the 24$\mu$m/0.7$\mu$m, 50$\mu$m/0.7$\mu$m and 100$\mu$m/0.7$\mu$m against 6$\mu$m/0.7$\mu$m ratios, for each combination of grain species and dust mass. These plots illustrate that for astronomical silicates and MgSiO$_3$, the model colors generally lie far from the observed values. Amorphous carbon is the only species that comes close to reproducing the observations, with the observed colors lying slightly above all model values for 0.001\,M$_\odot$, and slightly below all model values for 0.01\,M$_\odot$.

\begin{figure}
\includegraphics[width=0.48\textwidth, clip = True, trim = 0 0 0 0]{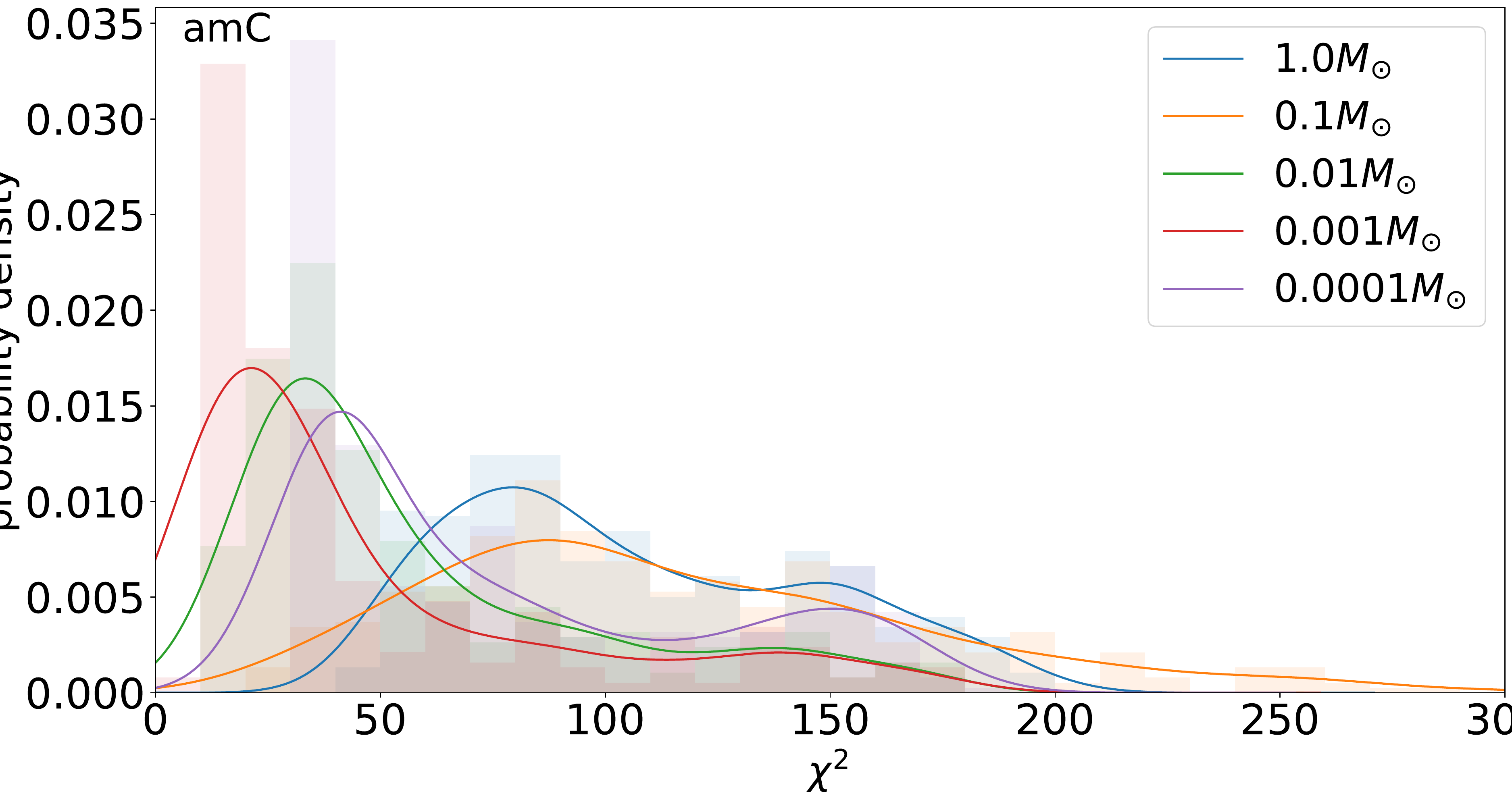}

\includegraphics[width=0.48\textwidth, clip = True, trim = 0 0 0 0]{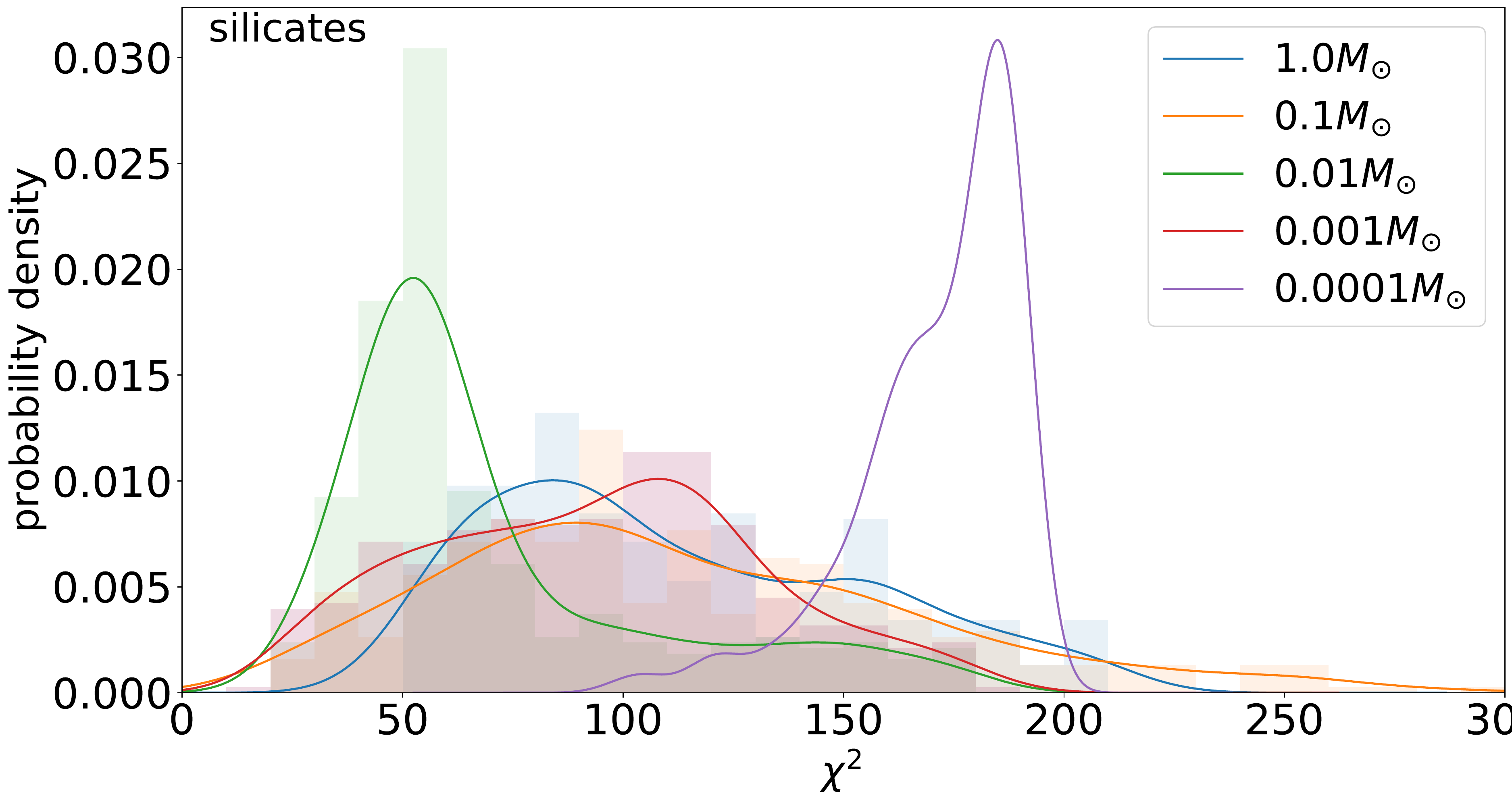}

\includegraphics[width=0.48\textwidth, clip = True, trim = 0 0 0 0]{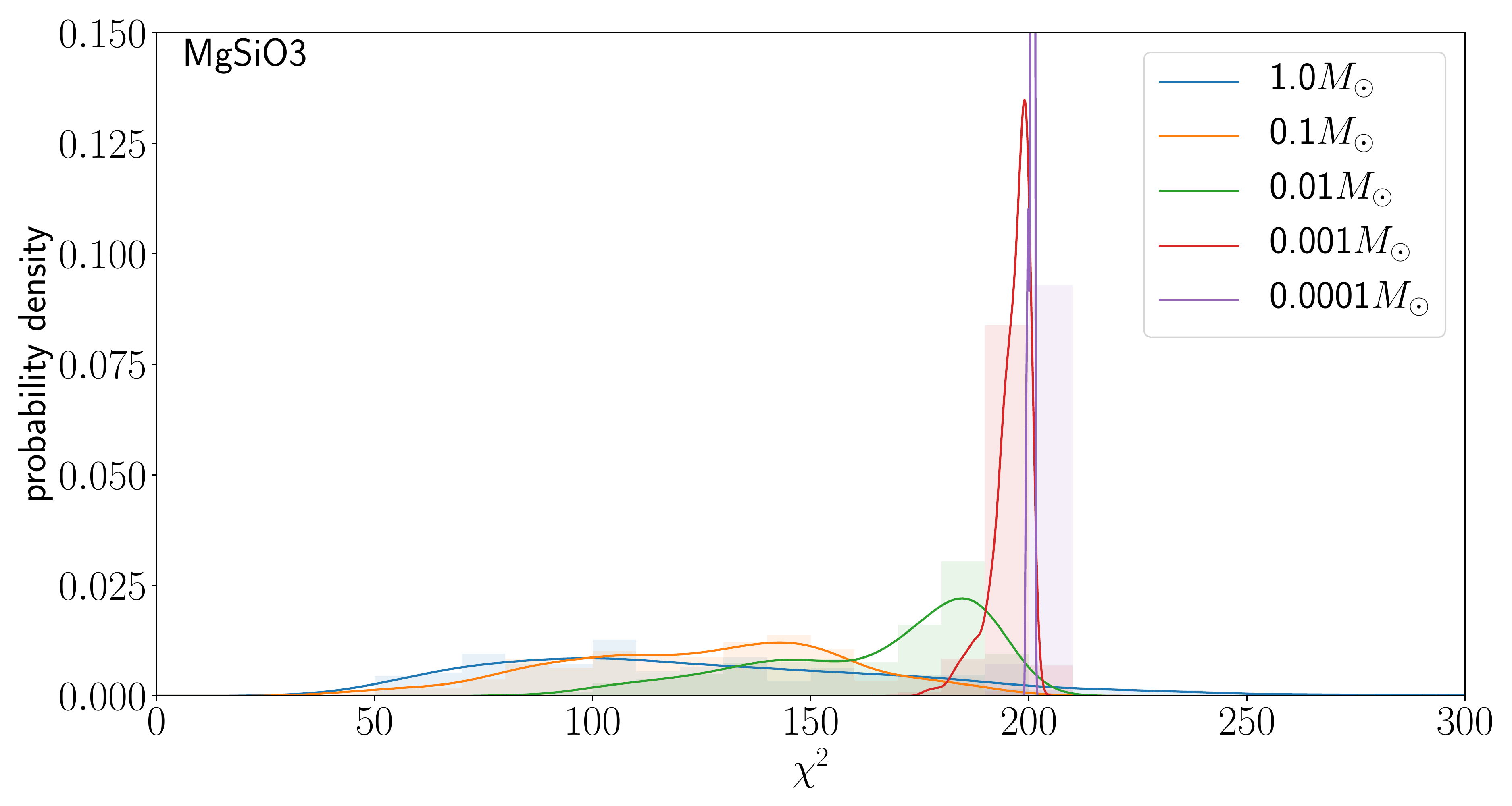}
\caption{Values of $\chi^2$ for all SED models in our parameter space for each of the dust masses investigated. It can be seen clearly that the vast majority of models result in poor fits; amorphous carbon models perform much better than either of the silicate species, and dust masses of 10$^{-3}$\,M$_\odot$ are the best-performing of the amorphous carbon set.}
\label{chi2_dust_mass}
\end{figure}

To quantify the goodness of fit, we calculated $\chi^2$ values by taking the average flux in the five regions of the spectrum, and comparing them to the equivalent average flux predicted by the models. We illustrate the variation of $\chi^2$ values for different dust masses and species in Figure \ref{chi2_dust_mass}. This figure shows that models containing 10$^{-3}$\,M$_\odot$ of amorphous carbon give the best fits to the SED. The best-fitting models with silicate grains have a dust mass of 10$^{-2}$\,M$_\odot$, because the lower opacity of silicate grains means that a much higher dust mass is required to fit the optical part of the SED, but these models are significantly worse than the lower-mass amorphous carbon models. MgSiO$_3$ performs worse still than the ``dirty" astronomical silicates; the lower opacity means that still higher masses are required to provide sufficient extinction in the optical, but the dust is too cool to adequately fit the IR part of the SED.

The performance of the whole suite of models is further discussed in Section~\ref{mocassin and damocles}. We identify in that section the regions of the parameter space that provide satisfactory fits to both the SED and the emission-line profiles.

\begin{figure*}
\centering
\includegraphics[height = 4.5cm, clip = True, trim = 20 0 100 0]{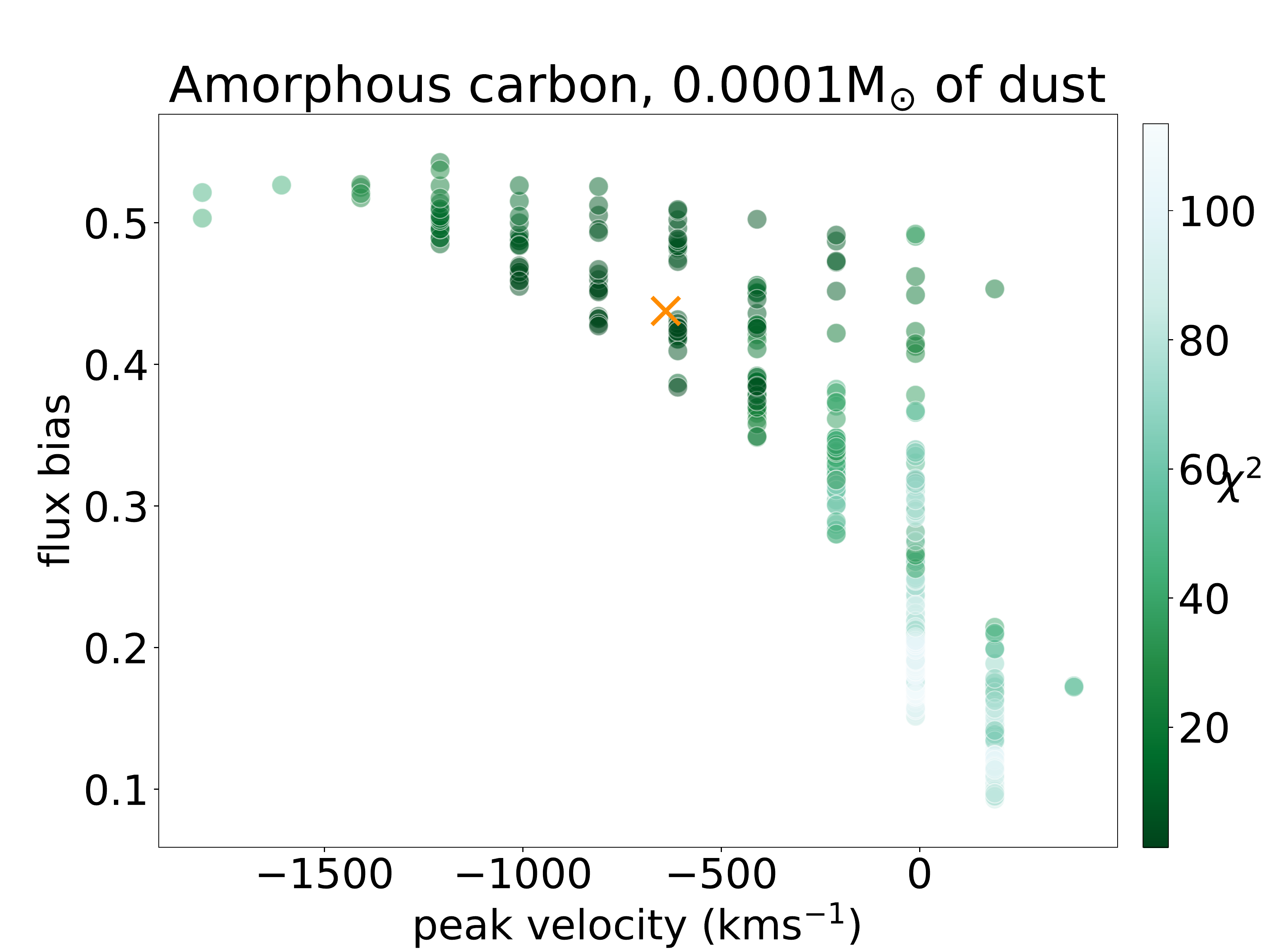}
\includegraphics[height = 4.5cm, clip = True, trim = 20 0 100 0]{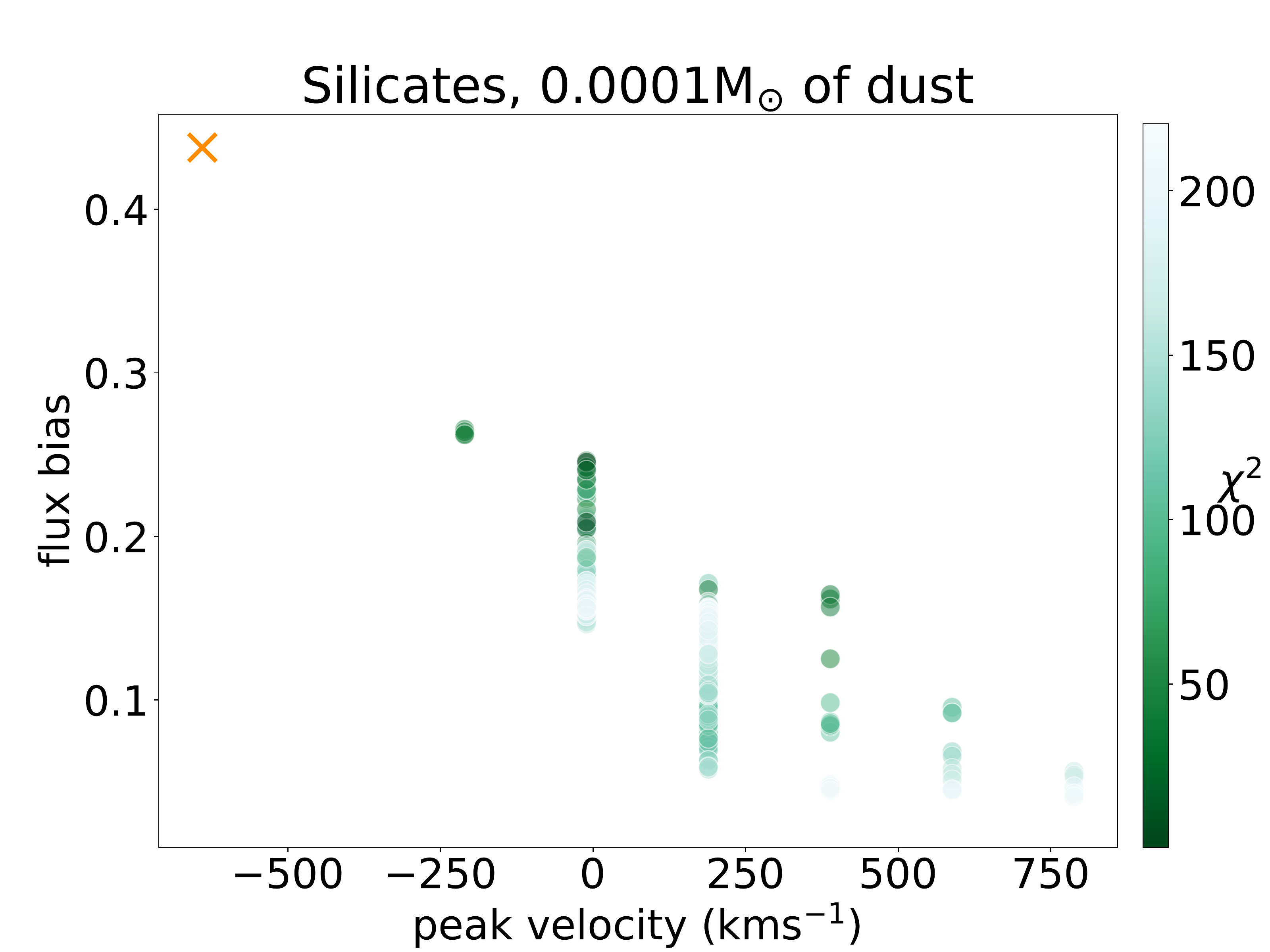}
\includegraphics[height = 4.5cm, clip = True, trim = 20 0 0 0]{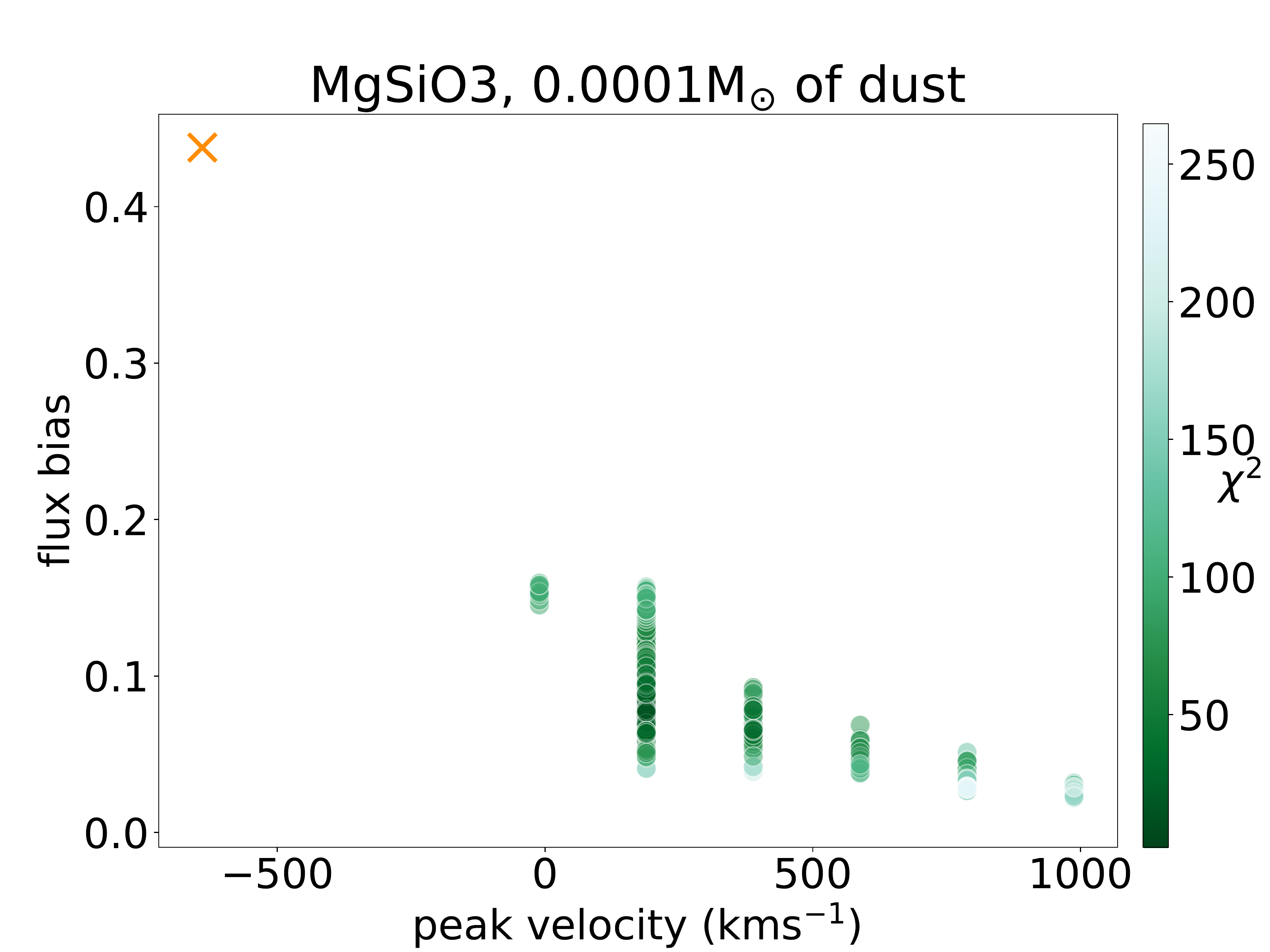}

\includegraphics[height = 4.5cm, clip = True, trim = 20 0 100 0]{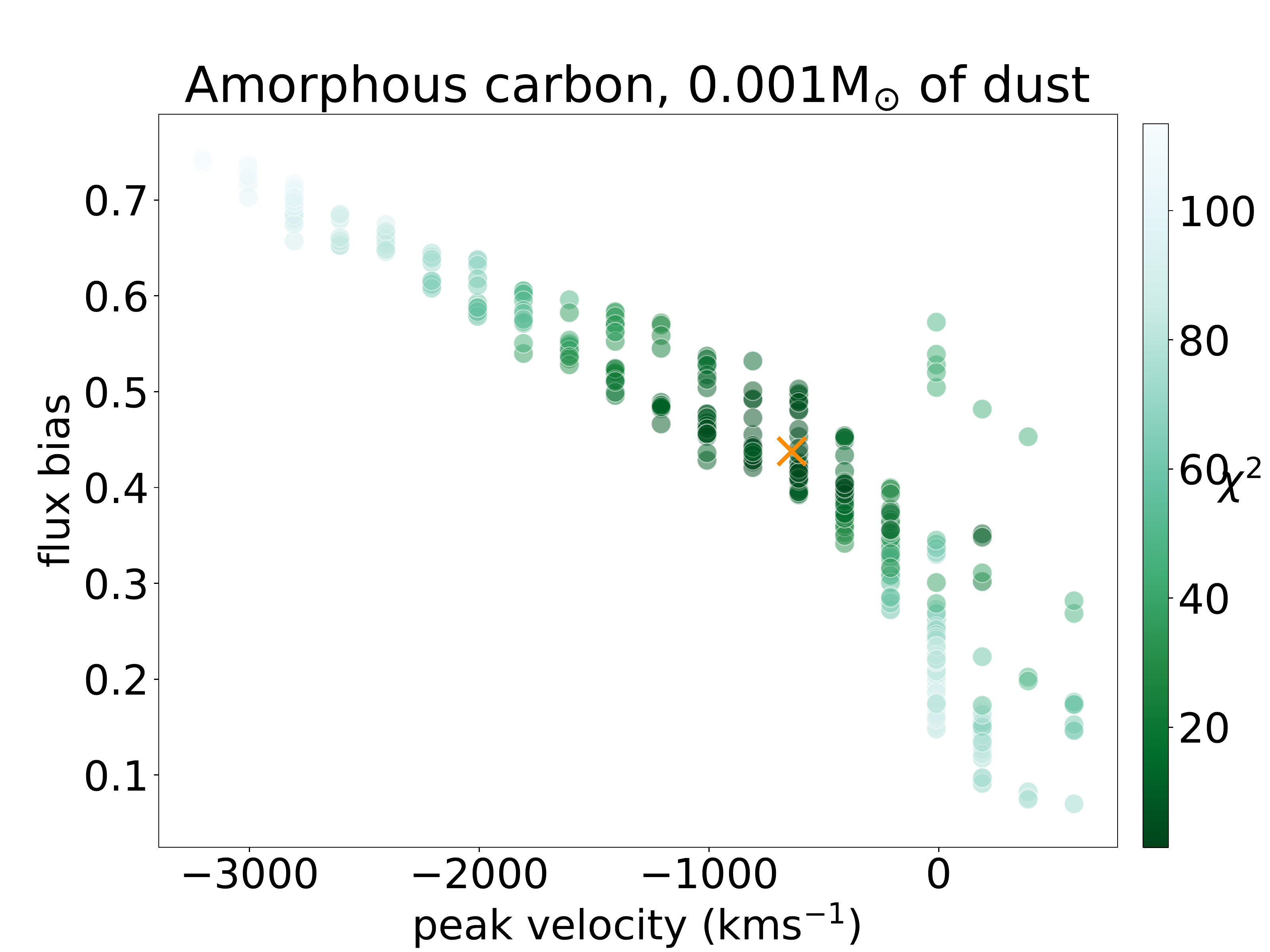}
\includegraphics[height = 4.5cm, clip = True, trim = 20 0 100 0]{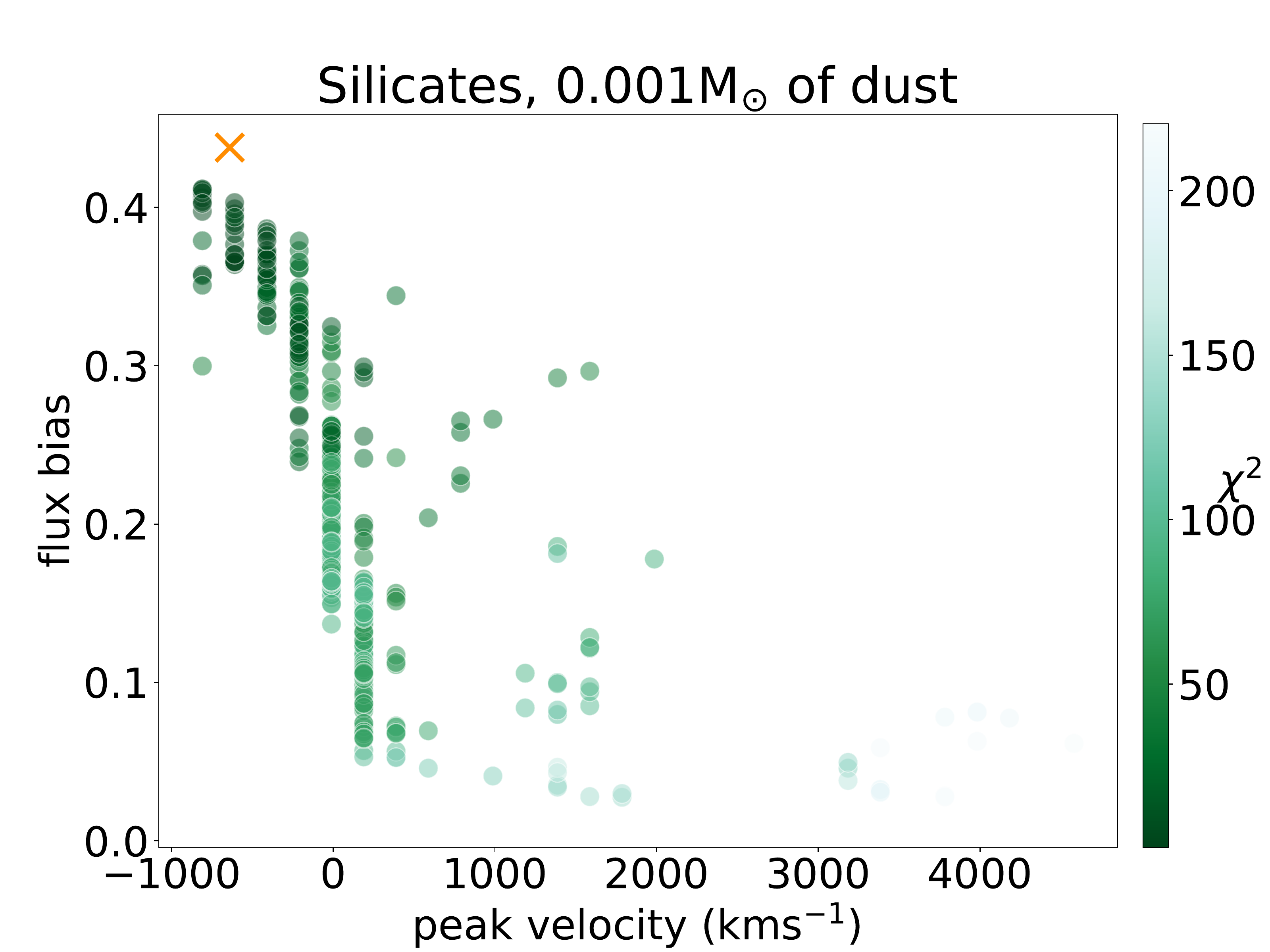}
\includegraphics[height = 4.5cm, clip = True, trim = 20 0 0 0]{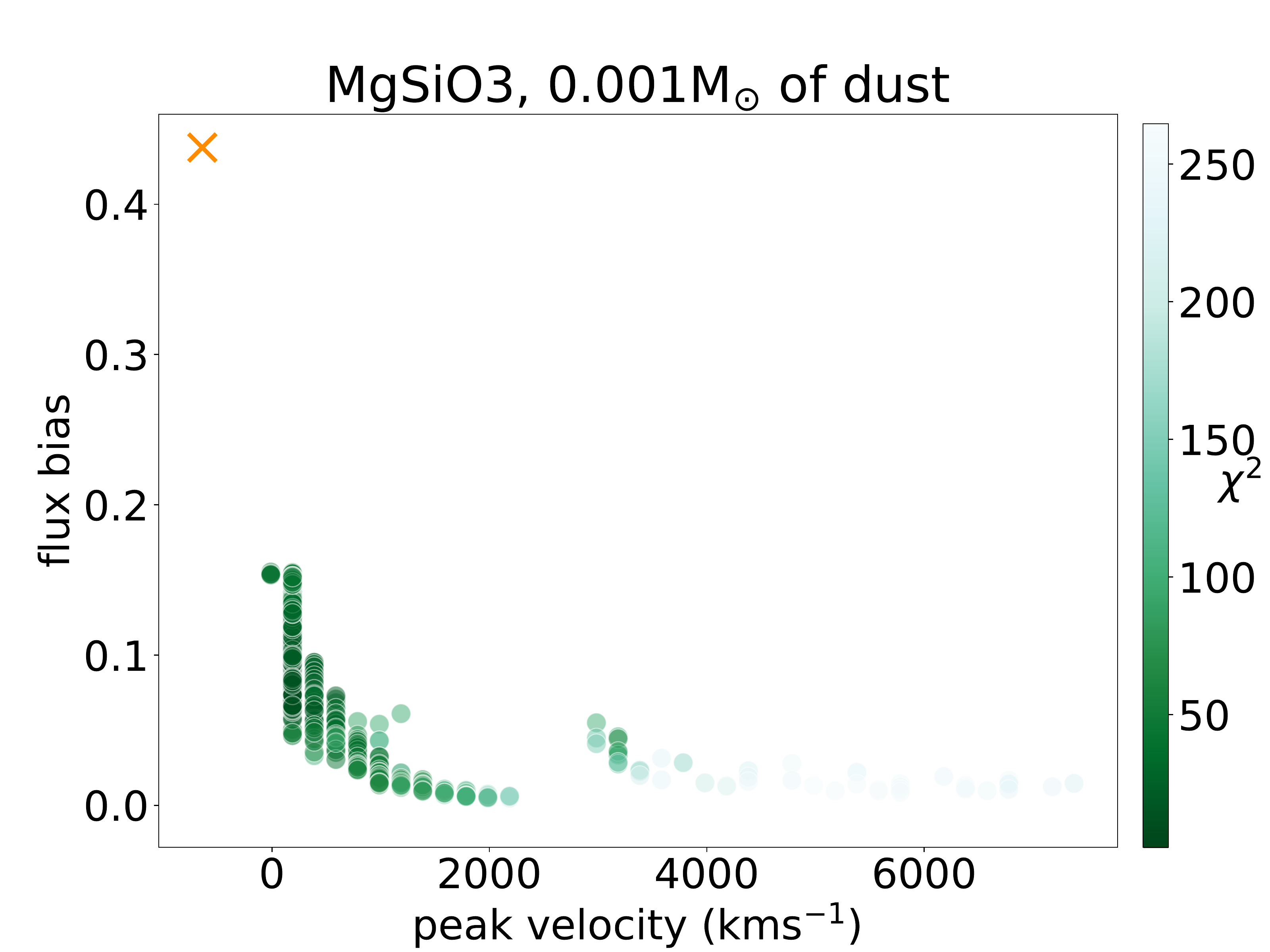}

\includegraphics[height = 4.5cm, clip = True, trim = 20 0 100 0]{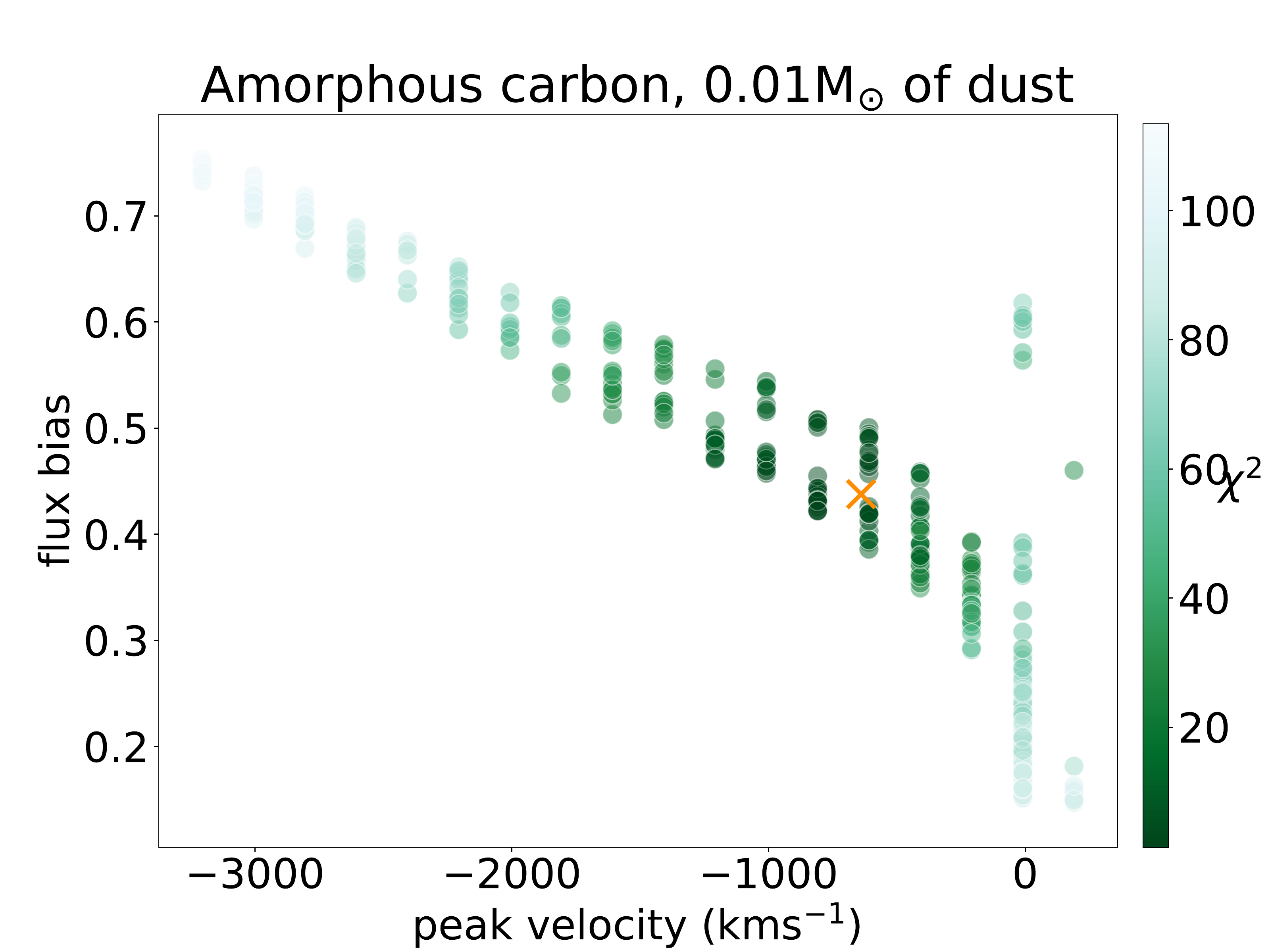}
\includegraphics[height = 4.5cm, clip = True, trim = 20 0 100 0]{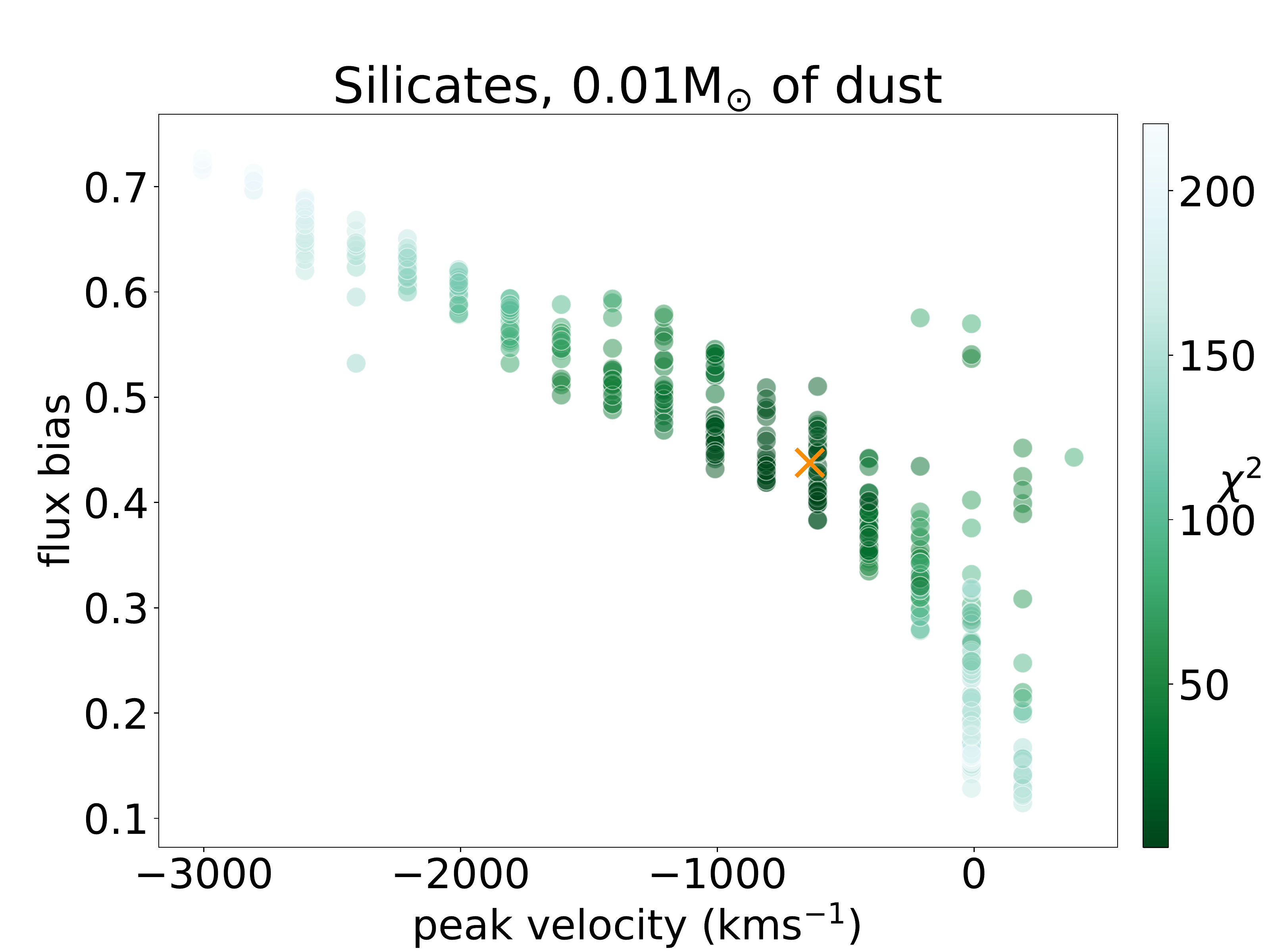}
\includegraphics[height = 4.5cm, clip = True, trim = 20 0 0 0]{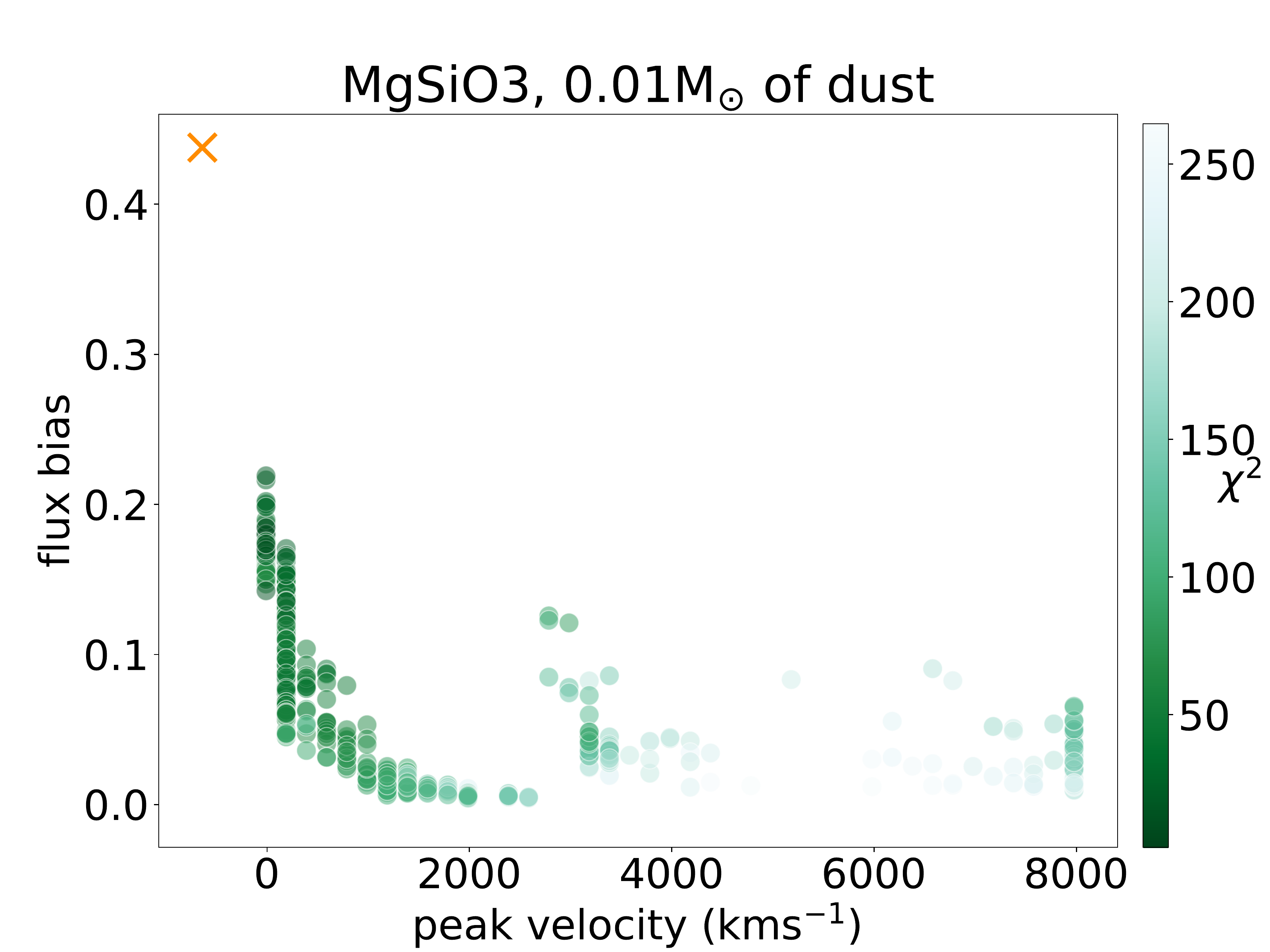}

\includegraphics[height = 4.5cm, clip = True, trim = 20 0 100 0]{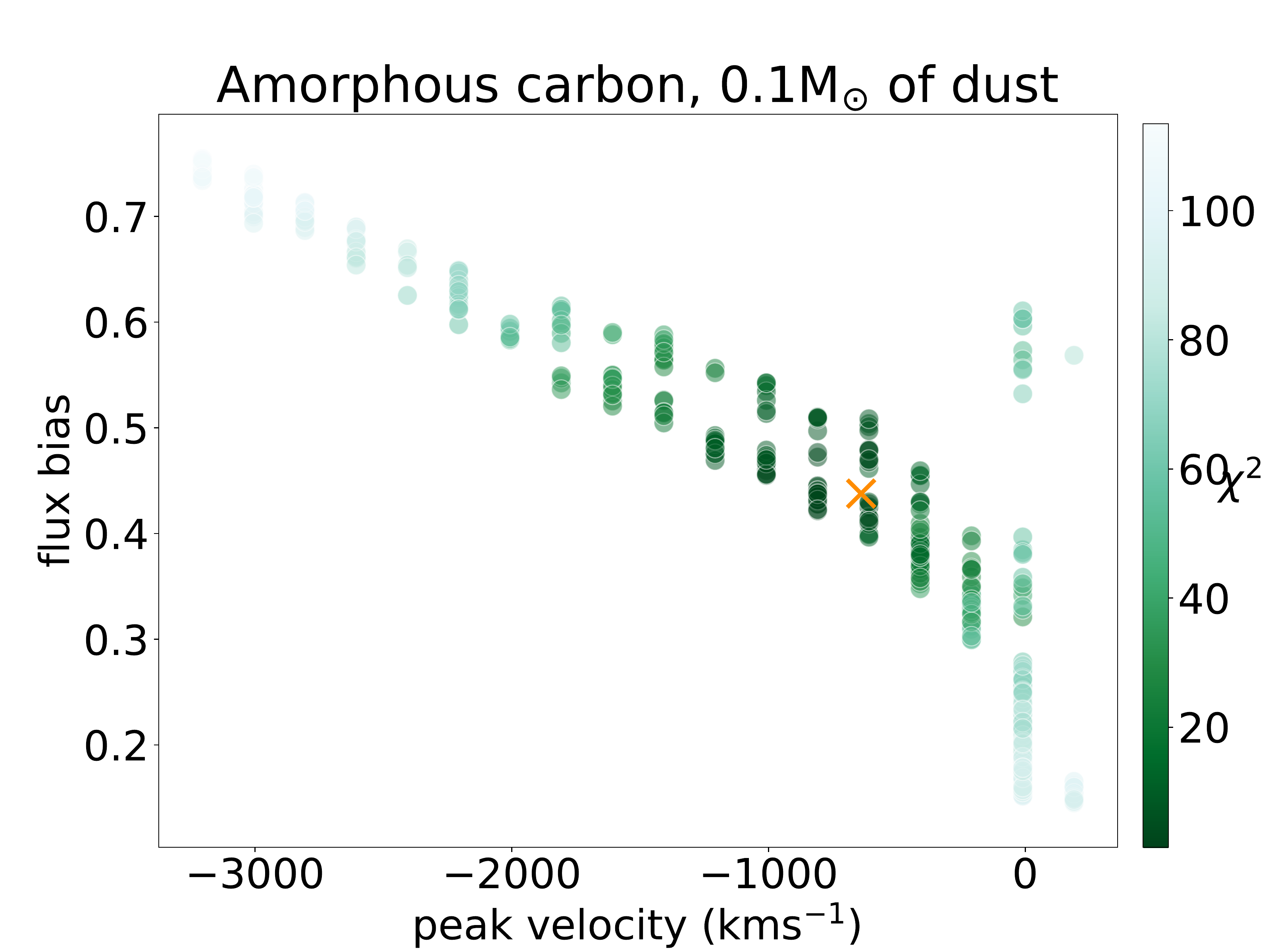}
\includegraphics[height = 4.5cm, clip = True, trim = 20 0 100 0]{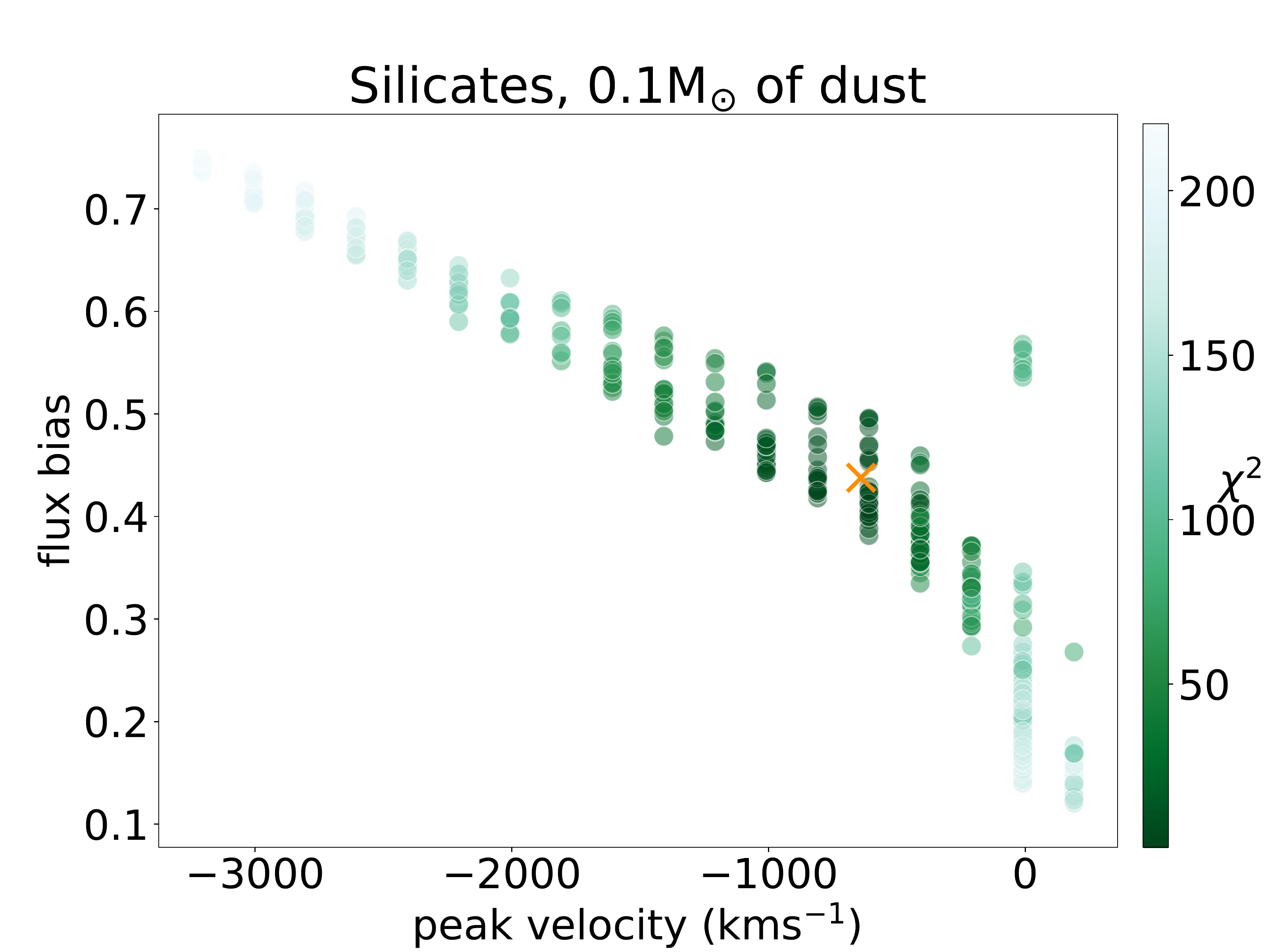}
\includegraphics[height = 4.5cm, clip = True, trim = 20 0 0 0]{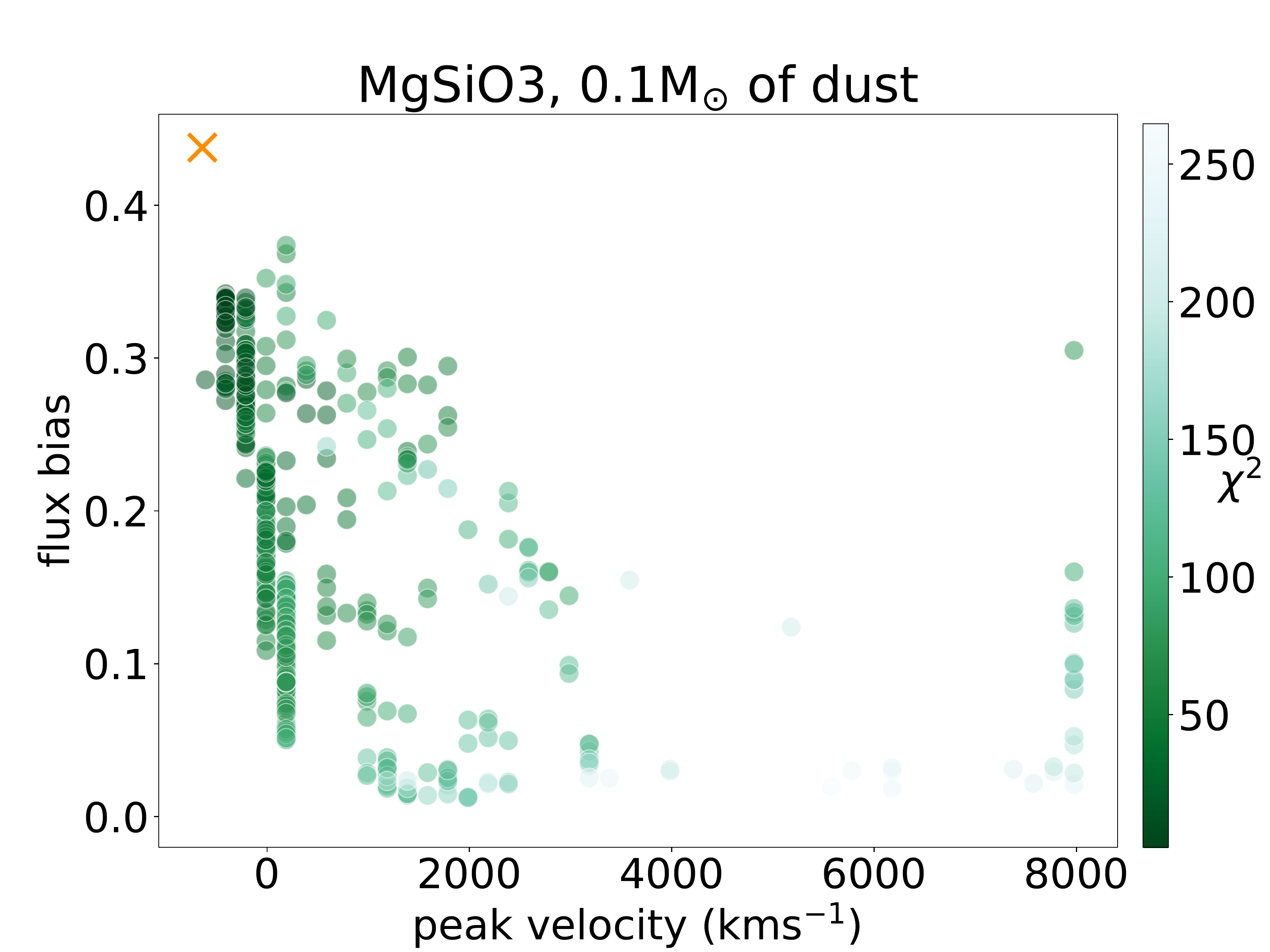}

\includegraphics[height = 4.5cm, clip = True, trim = 20 0 100 0]{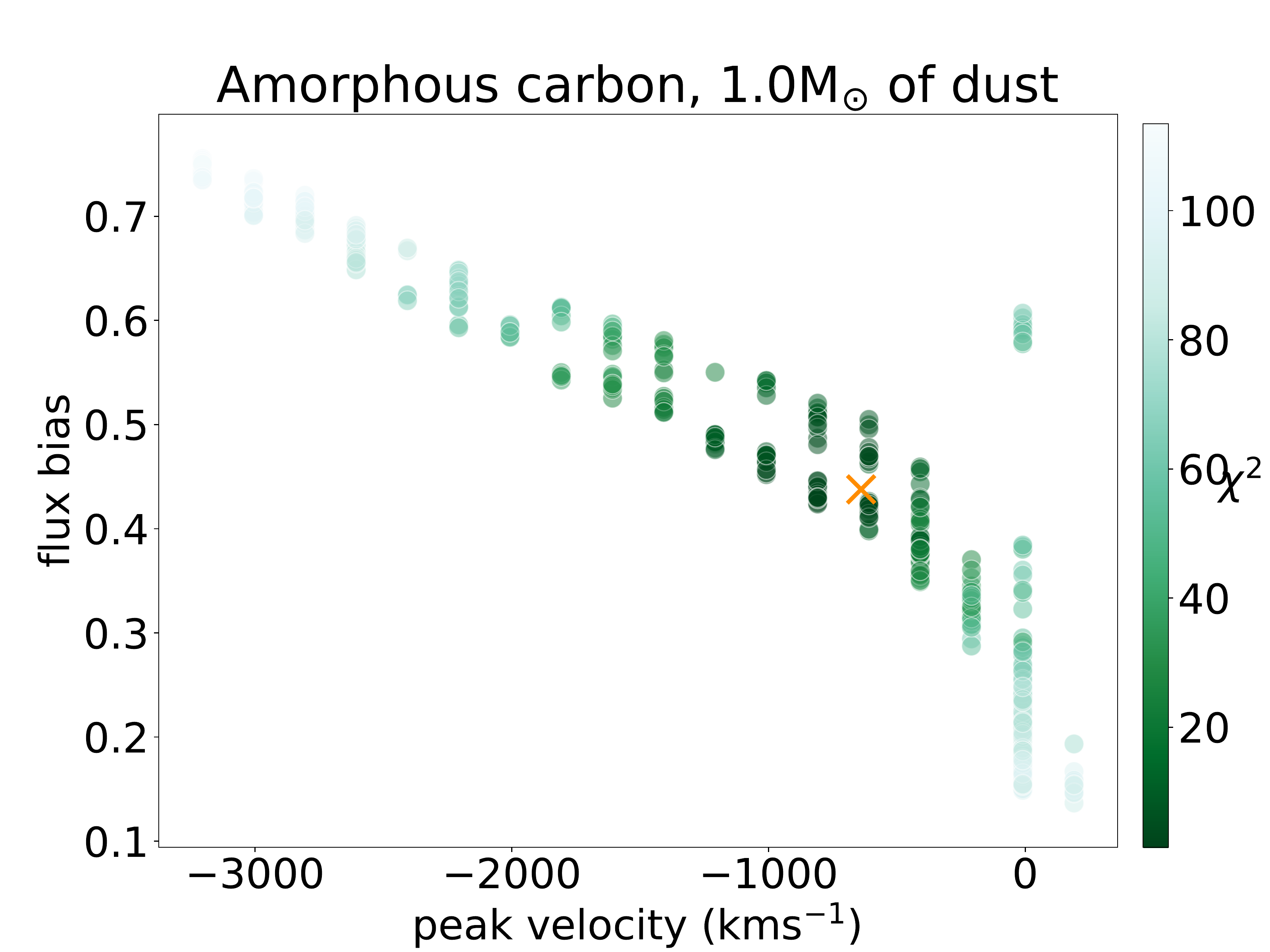}
\includegraphics[height = 4.5cm, clip = True, trim = 20 0 100 0]{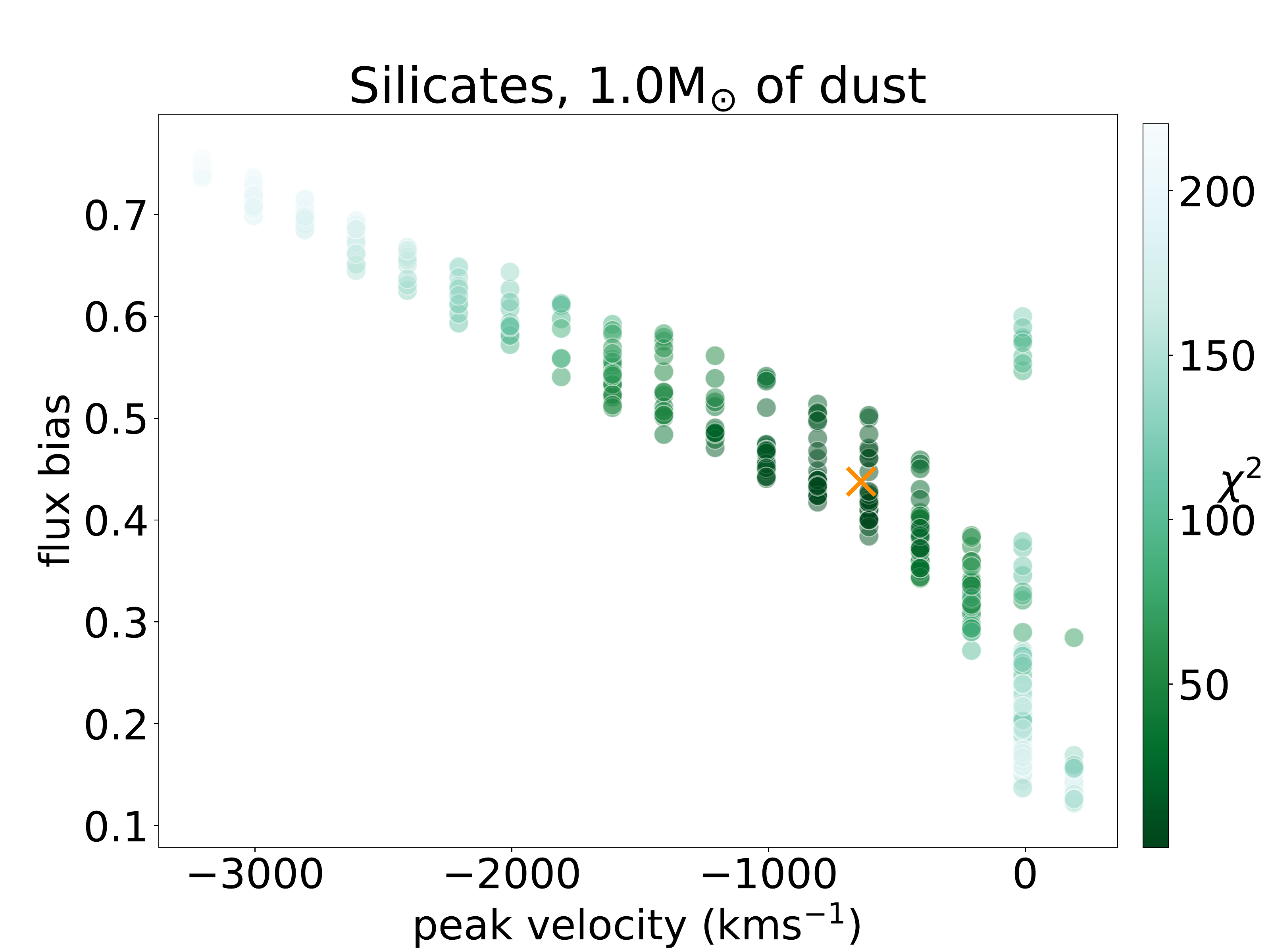}
\includegraphics[height = 4.5cm, clip = True, trim = 20 0 0 0]{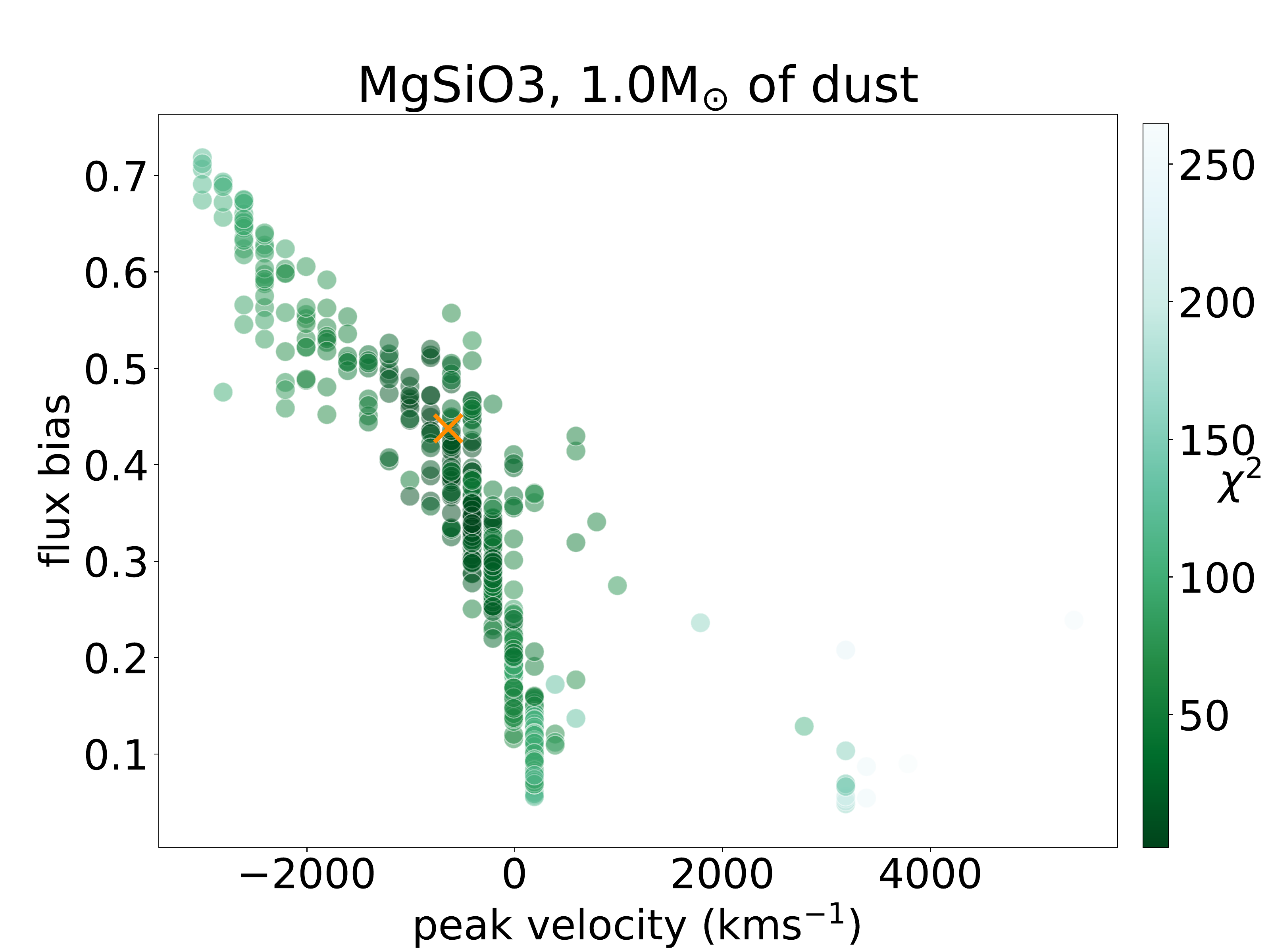}

\caption{Velocity of peak flux against overall flux bias (the fraction of the flux on the blue side of the profile) for all [O~{\sc i}]~6300,6363~\AA\ models. The value of $\chi^2$ is shown as variations in the color of the points. The observed values for the [O~{\sc i}]~6300,6363~\AA\ doublet of SN~1987A at 806 days post-outburst is shown as an orange cross. As expected, models with the lowest $\chi^2$ are closest to the observed values.}
\label{chi2_maps}
\end{figure*}

\begin{figure}
\centering
\includegraphics[width=0.42\textwidth, clip = True, trim = 10 20 38 40]{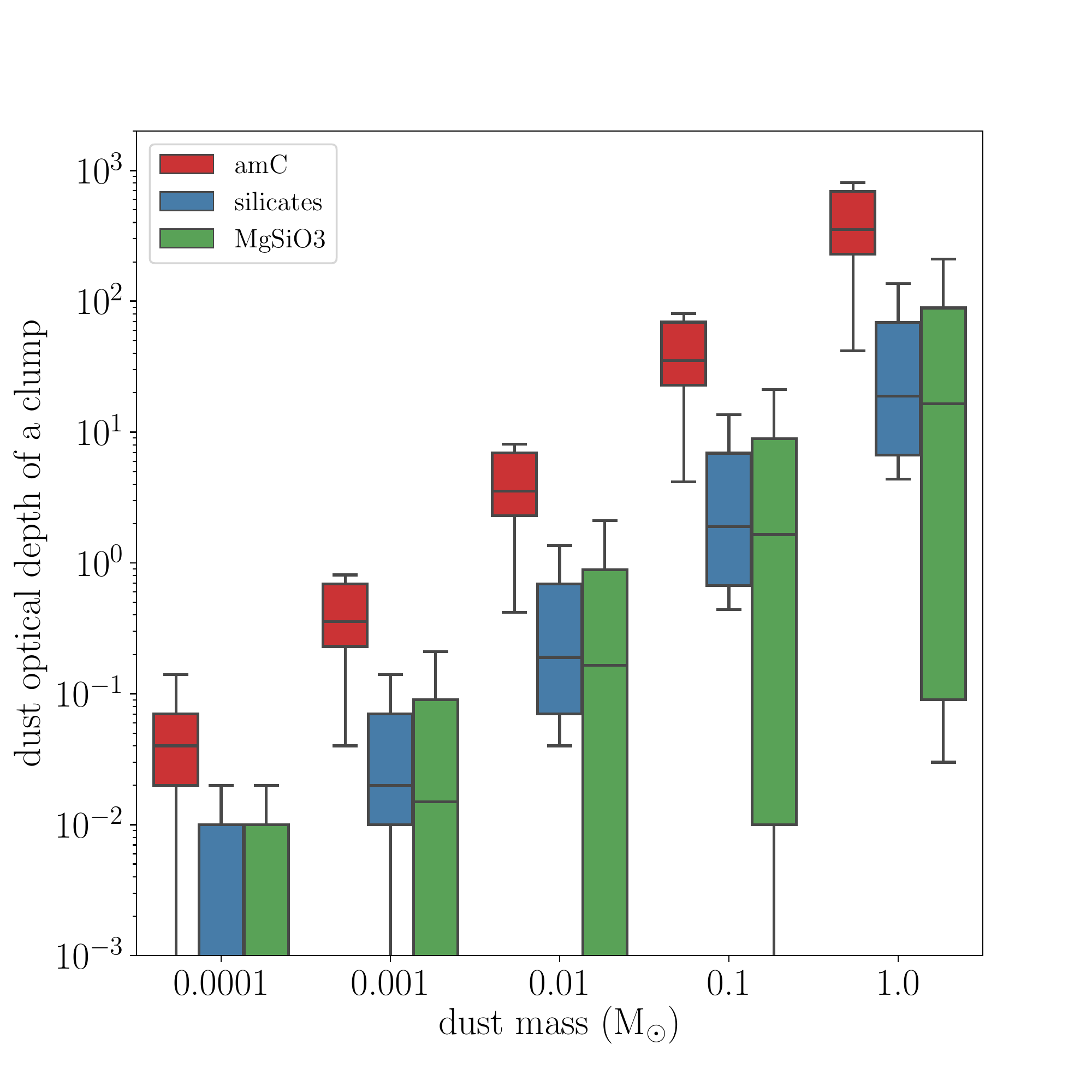}
\caption{Box plots showing the range of clump dust optical depths at H$\alpha$ for different dust masses and species for the emission line profile models. The spread in values is due to variations in clump size, grain radius and volume filling factor.}
\label{fig_clump_optical_depths}
\end{figure}

\begin{figure}
\centering
\includegraphics[width=0.4\textwidth, clip = True, trim = 10 0 38 0]{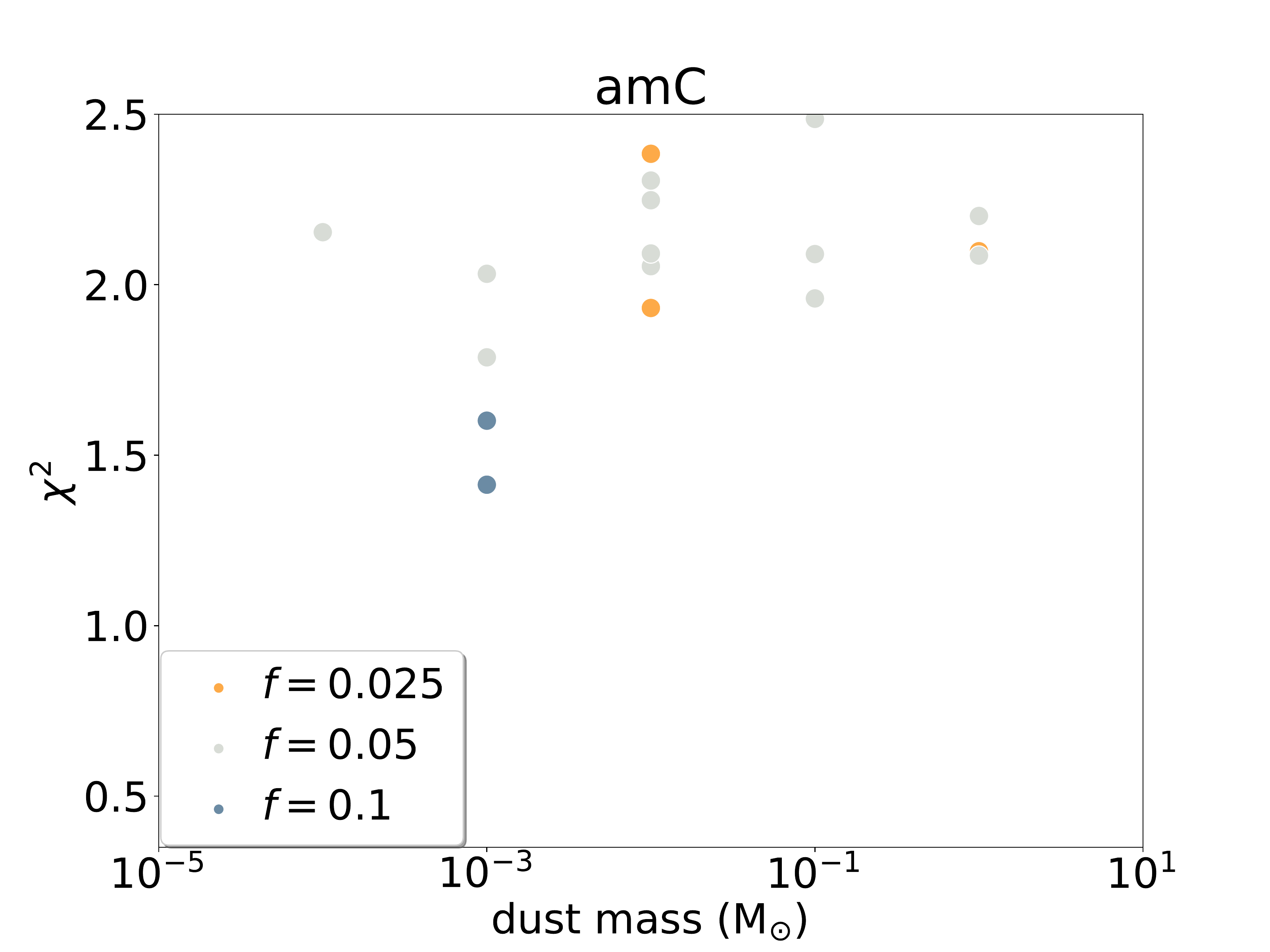}
\includegraphics[width=0.4\textwidth, clip = True, trim = 10 0 38 0]{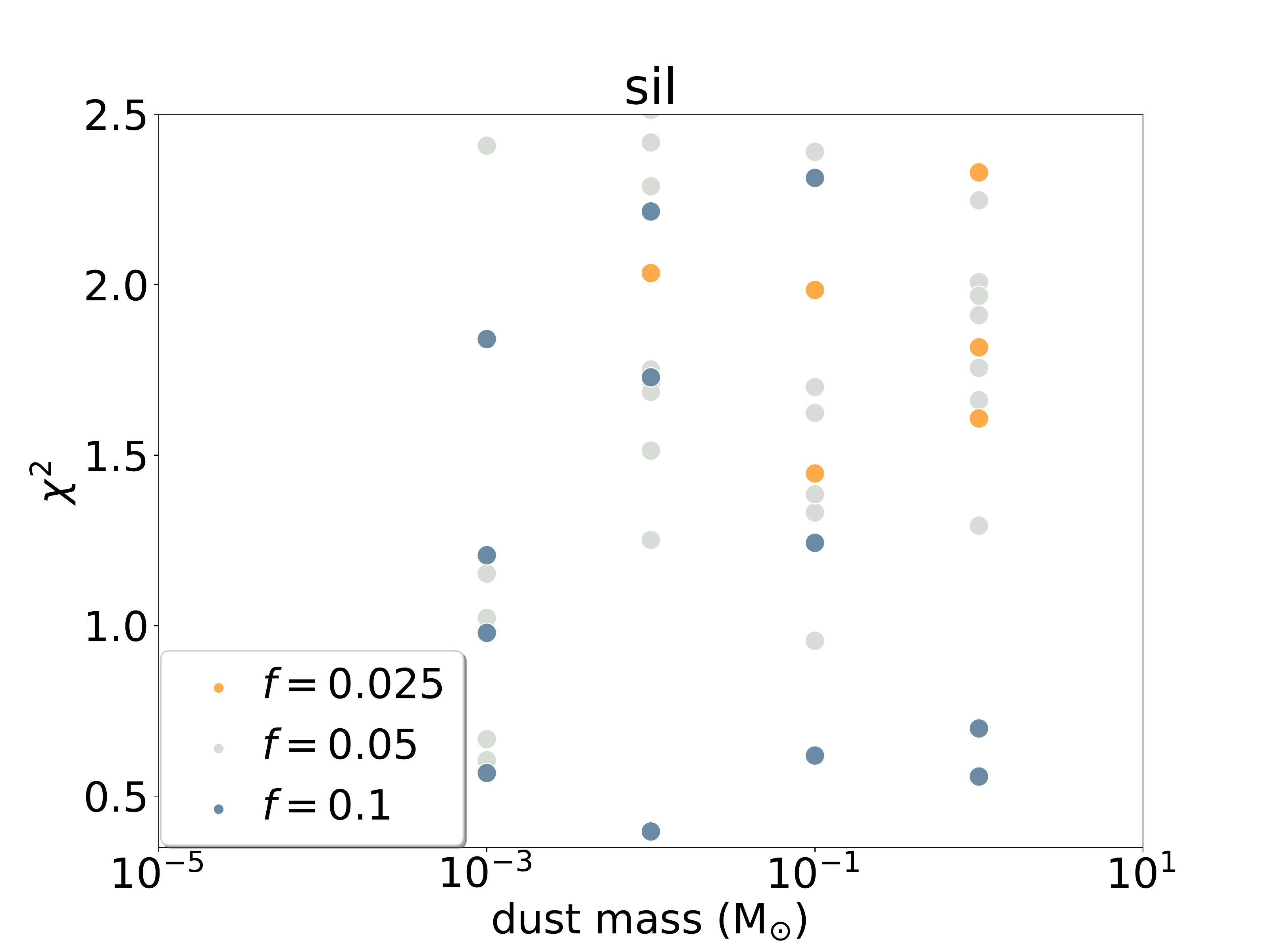}
\includegraphics[width=0.4\textwidth, clip = True, trim = 10 0 38 0]{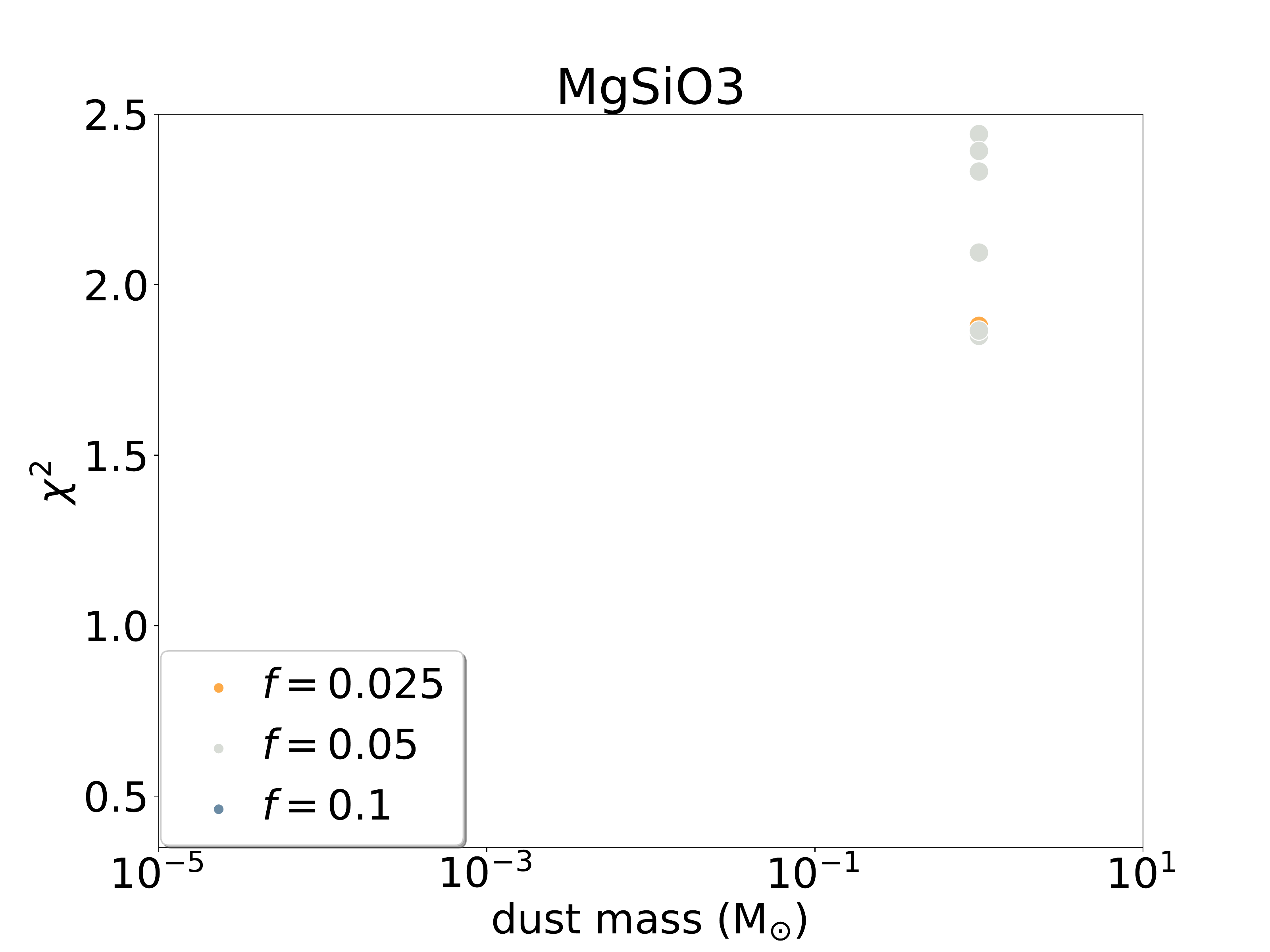}

\caption{The variation of $\chi^2$ with dust mass for the best-fitting ($\chi^2<2.5$) dust emission line profile models for different dust species. The filling factors required for each model are shown in different colors.}
\label{Mdust_vs_chi2}
\end{figure}

\subsection{Comparison of emission line profile models to observations}

To compare our grid of emission line profile models to the observed [O~{\sc i}]~6300,6363~\AA\ doublet at 806 days, we convolved the modeled profiles to the spectral resolution of the observed spectrum (16\AA) and scaled them to match the peak flux of the observed [O~{\sc i}]~6300,6363~\AA\ doublet. A number of metrics were calculated including the overall flux bias of the doublet (i.e. the fraction of the flux on the blue side of the profile), the velocity of the peak flux and $\chi^2$ calculated across the width of the doublet.

In Figure \ref{chi2_maps}, we show results for all of the emission line models in the grid. We calculated the velocity of the peak flux and the overall flux bias (note that this was calculated for the entire doublet, not for one component). We also show that the variation of $\chi^2$ was as expected, with regions of highest likelihood corresponding to those models with flux biases and peak fluxes nearest the observed values. We mark on these plots the true values for the observed [O~{\sc i}]~6300,6363~\AA\ doublet. The majority of models resulted in poor fits. However, at least a few reasonable fits to the profile can be obtained for all dust species. Amorphous carbon grains can reproduce the observed profile for any dust mass if specific values for the other parameters are selected, whilst silicates and MgSiO$_3$ grains fail to reproduce the line profile for smaller dust masses under any circumstances. This is due to the glassy nature of silicate and MgSiO$_3$ grains, which are significantly more scattering that carbonaceous grains. Consequently, for low dust masses and therefore optically thin dust configurations, the modeled [O~{\sc i}]~6300,6363~\AA\ doublet becomes significantly redshifted due to repeated scattering events experienced by photons doing work on the expanding ejecta. For higher dust masses, the optical depths are increased which reduces the opportunity for repeated scattering events and thus the profiles are more symmetrical or blueshifted resulting in improved fits to the observed profile. The variation in clump dust optical depths over the grid is presented in Figure \ref{fig_clump_optical_depths}.

\begin{figure*}
\includegraphics[width=\textwidth, clip = True, trim = 0 0 0 0]{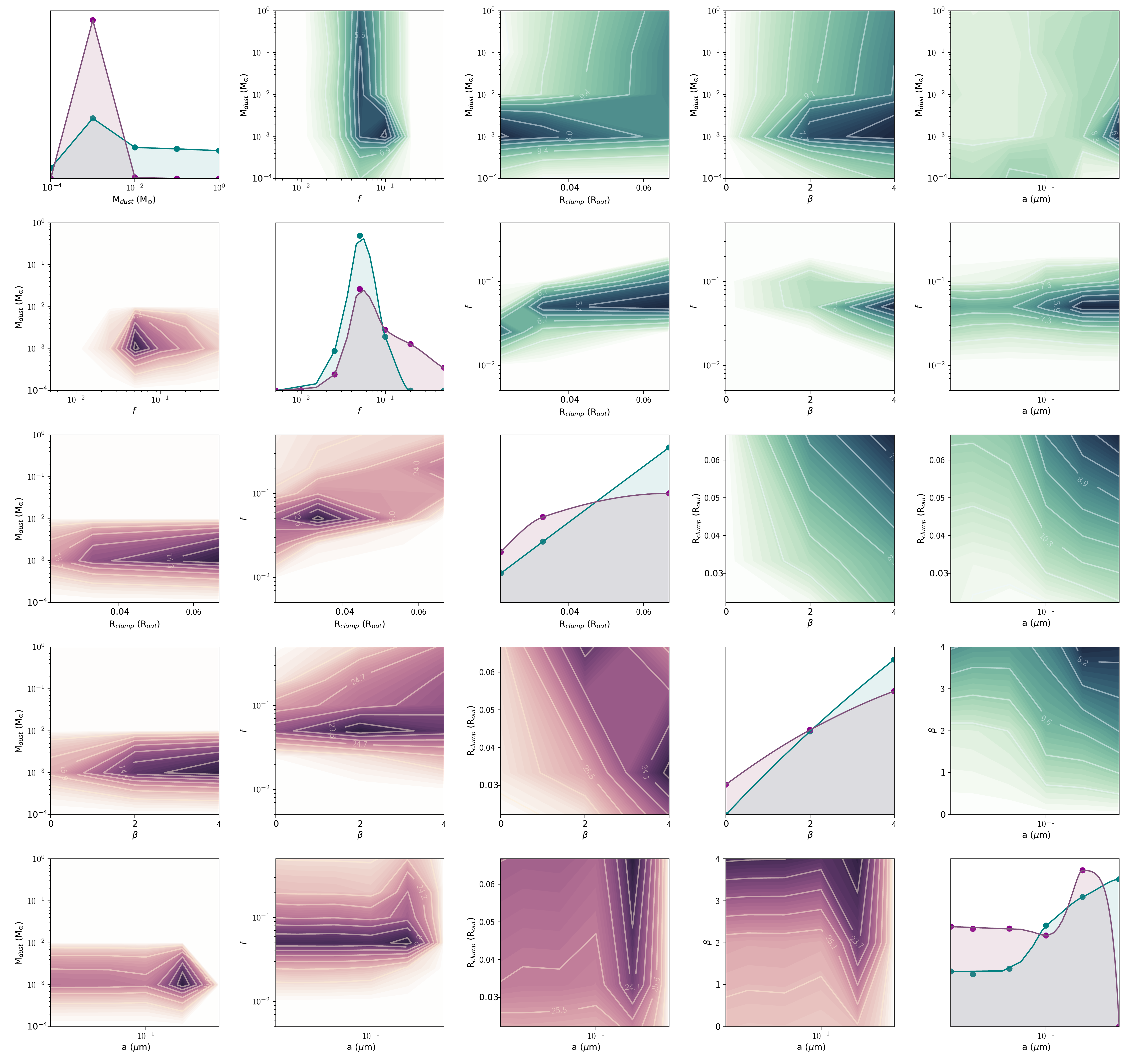}
\caption{2D normalized likelihood contours for the amorphous carbon grain line profile models are presented in green (top half) and for the SED models in purple (lower half), marginalized over the other parameters. Corresponding $\chi^2$ values are quoted on the plots for specific contours. 1D normalized likelihood distributions are given for each parameter on the diagonal with the SED 1D likelihoods plotted in purple and the line profile 1D likelihoods plotted in green. }
\label{corner_amC}
\end{figure*}

\begin{figure*}
\includegraphics[width=\textwidth, clip = True, trim = 0 0 0 0]{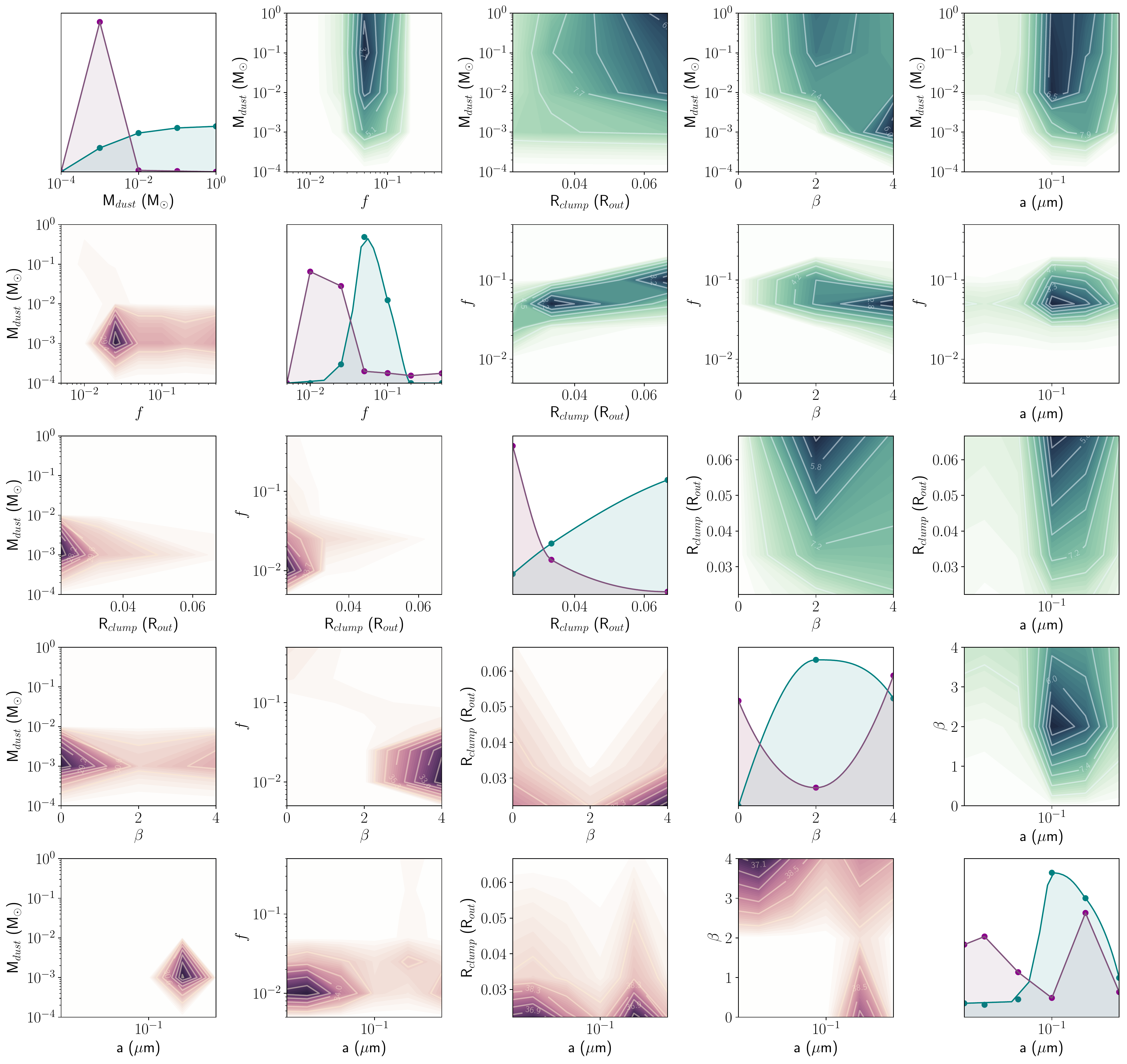}
\caption{2D normalized likelihood contours for the silicate grain line profile models are presented in green (top half) and for the SED models in purple (lower half), marginalized over the other parameters. Corresponding $\chi^2$ values are quoted on the plots for specific contours. 1D normalized likelihood distributions are given for each parameter on the diagonal with the SED 1D likelihoods plotted in purple and the line profile 1D likelihoods plotted in green. }
\label{corner_sil}
\end{figure*}

\begin{figure*}
\includegraphics[width=\textwidth, clip = True, trim = 0 0 0 0]{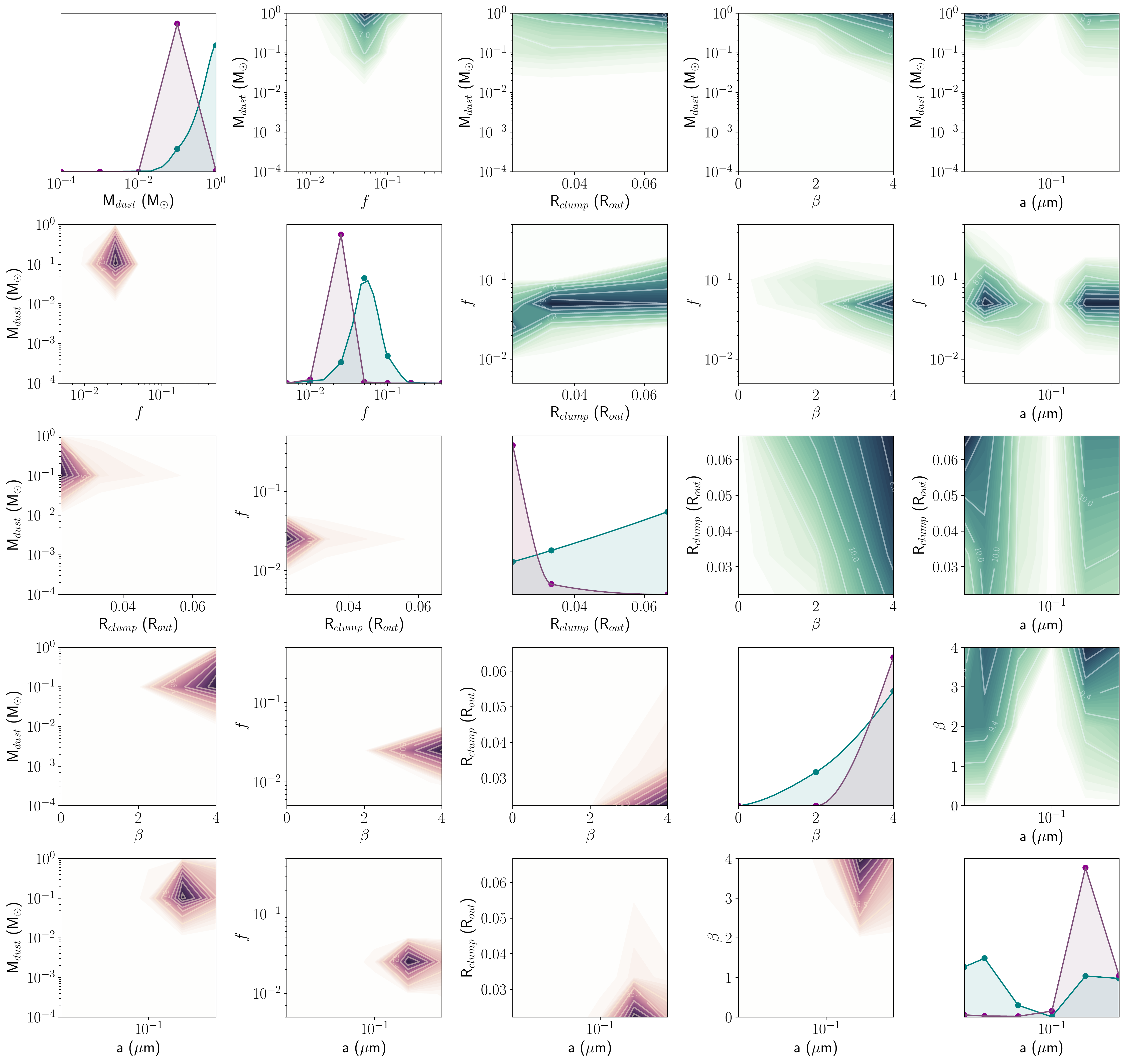}
\caption{2D normalized likelihood contours for the MgSiO$_3$ grain line profile models are presented in green (top half) and for the SED models in purple (lower half), marginalized over the other parameters. Corresponding $\chi^2$ values are quoted on the plots for specific contours. 1D normalized likelihood distributions are given for each parameter on the diagonal with the SED 1D likelihoods plotted in purple and the line profile 1D likelihoods plotted in green. }
\label{corner_MgSiO3}
\end{figure*}

\begin{figure*}
\includegraphics[width=0.32\textwidth, clip = True, trim = 0 0 0 0]{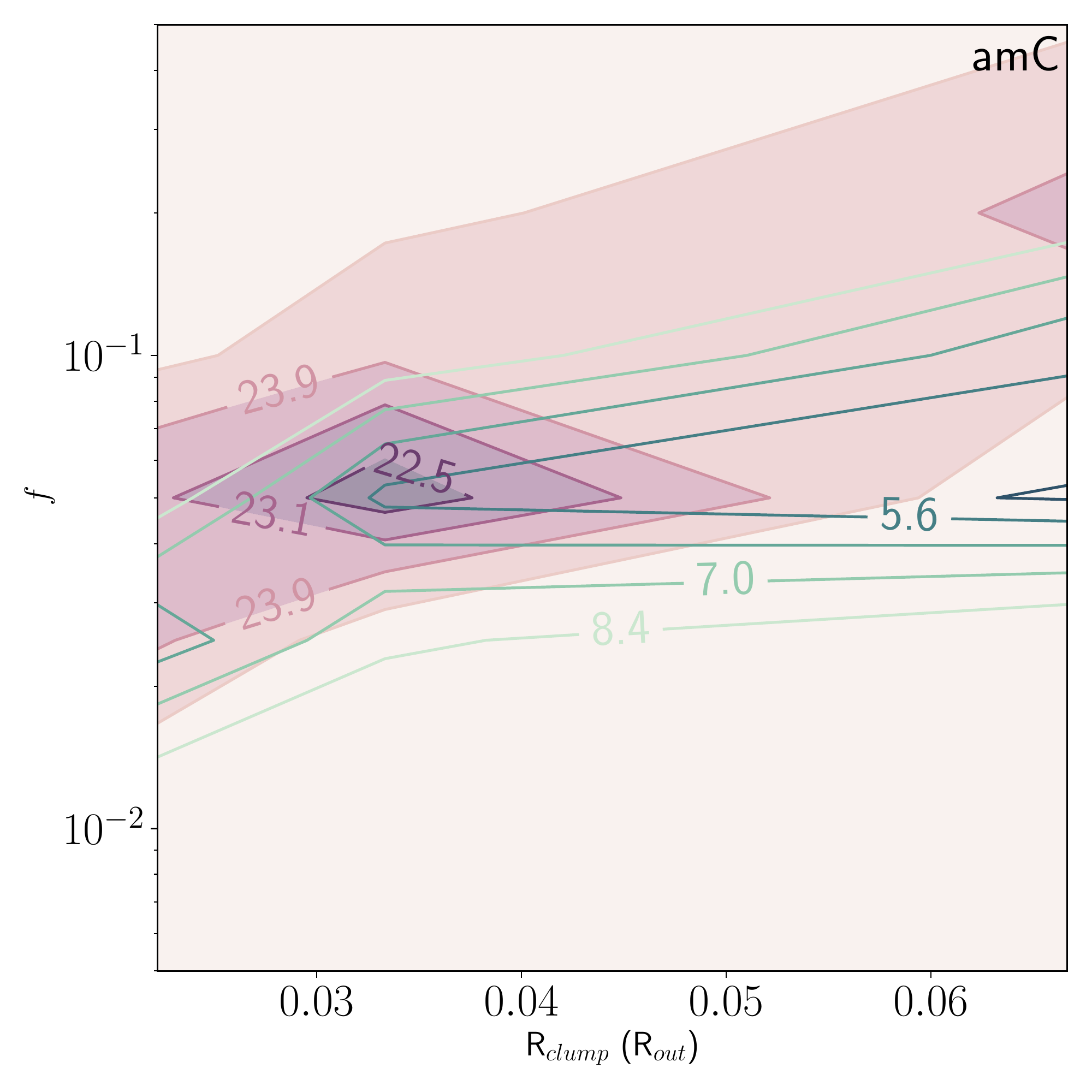}
\includegraphics[width=0.32\textwidth, clip = True, trim = 0 0 0 0]{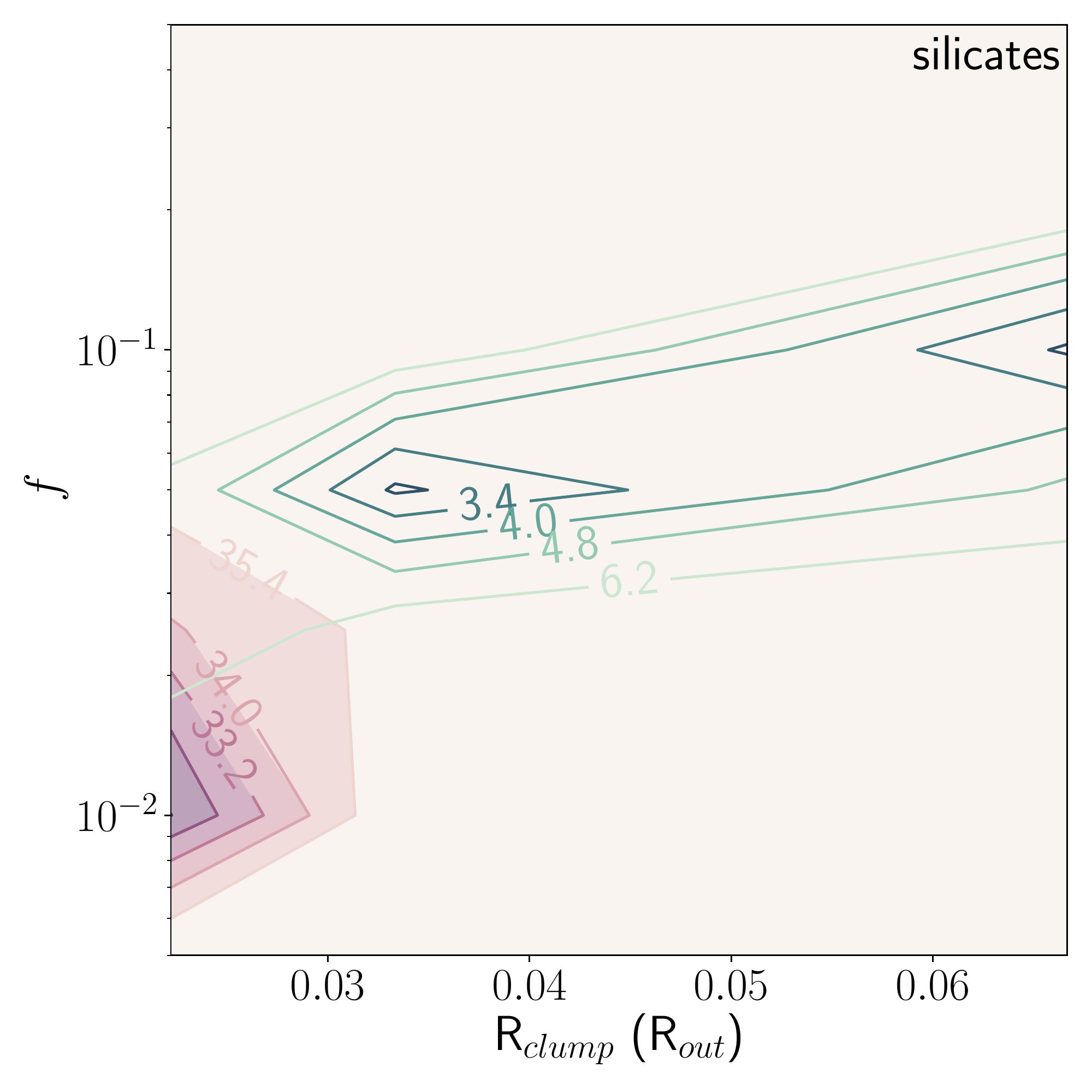}
\includegraphics[width=0.32\textwidth, clip = True, trim = 0 0 0 0]{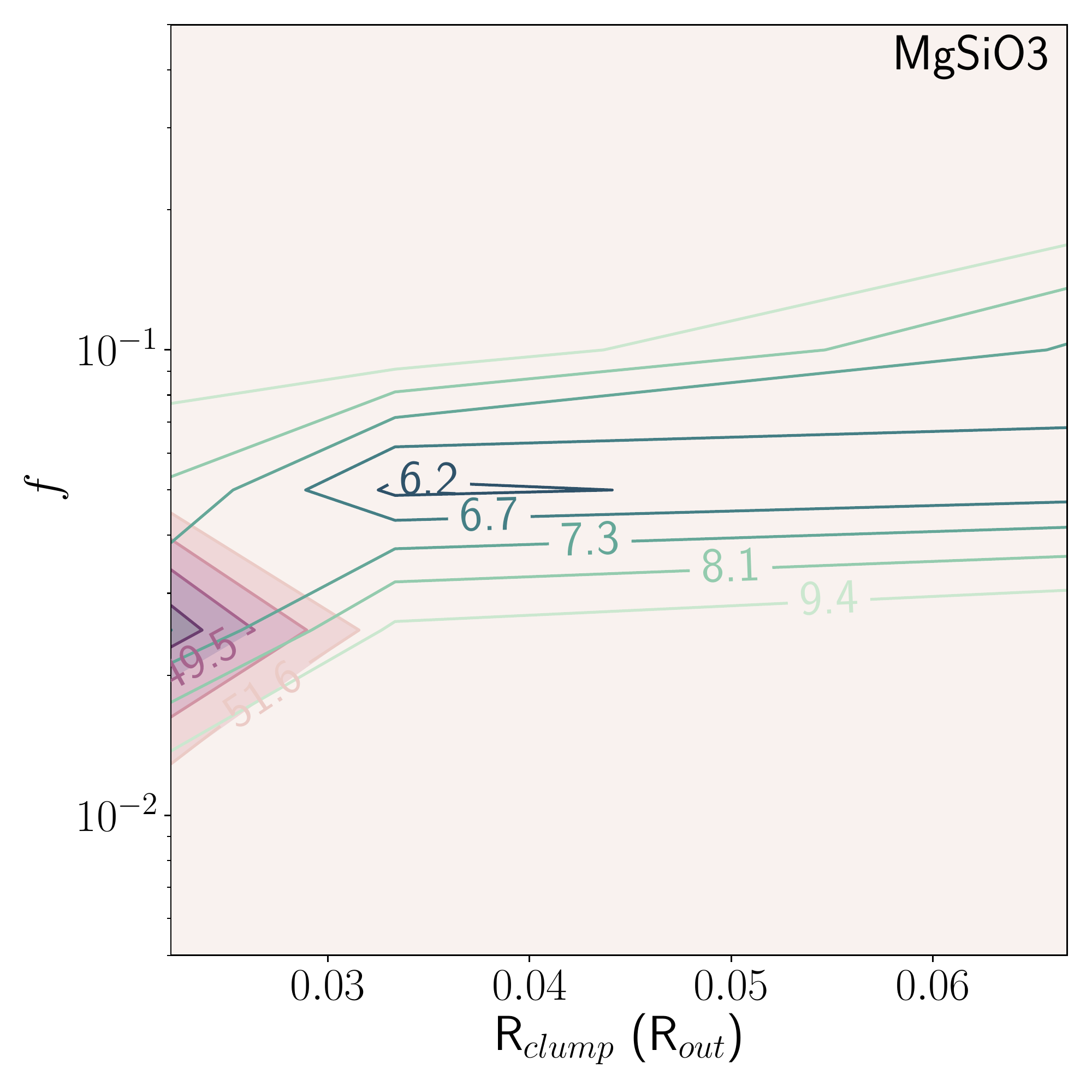}

\hspace{1pt}
\includegraphics[width=0.32\textwidth, clip = True, trim = 10 0 10 0]{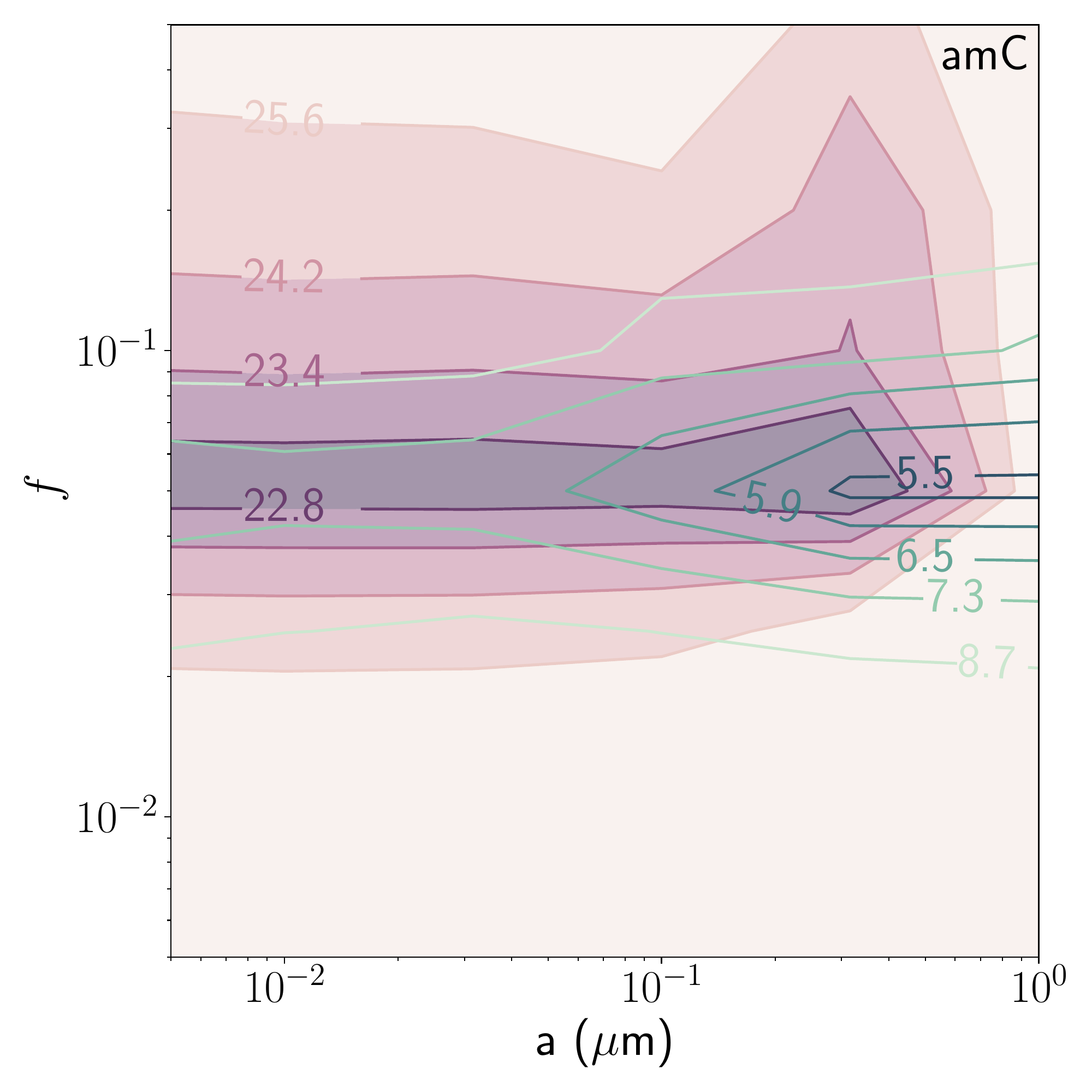}
\includegraphics[width=0.32\textwidth, clip = True, trim = 10 0 10 0]{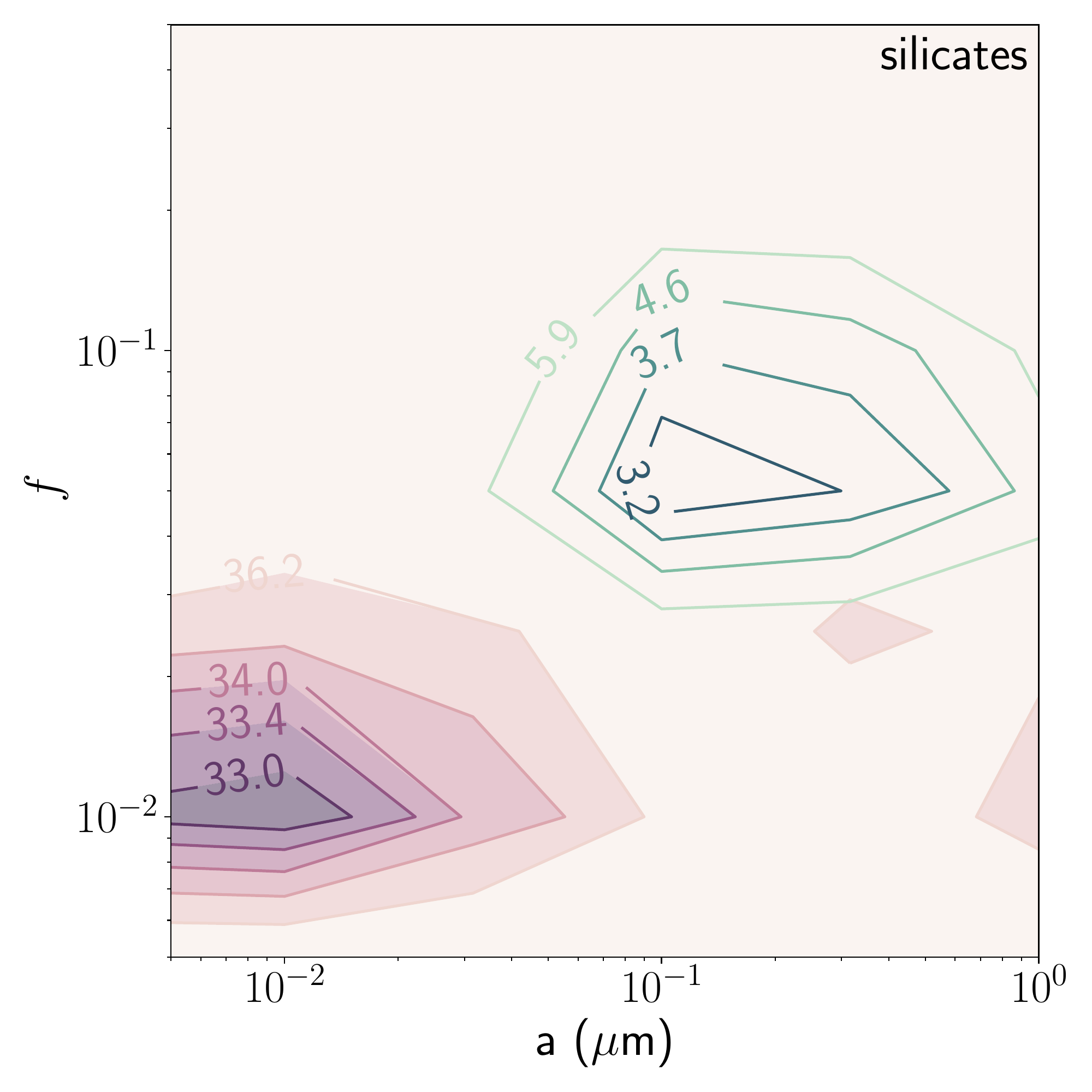}
\includegraphics[width=0.32\textwidth, clip = True, trim = 10 0 10 0]{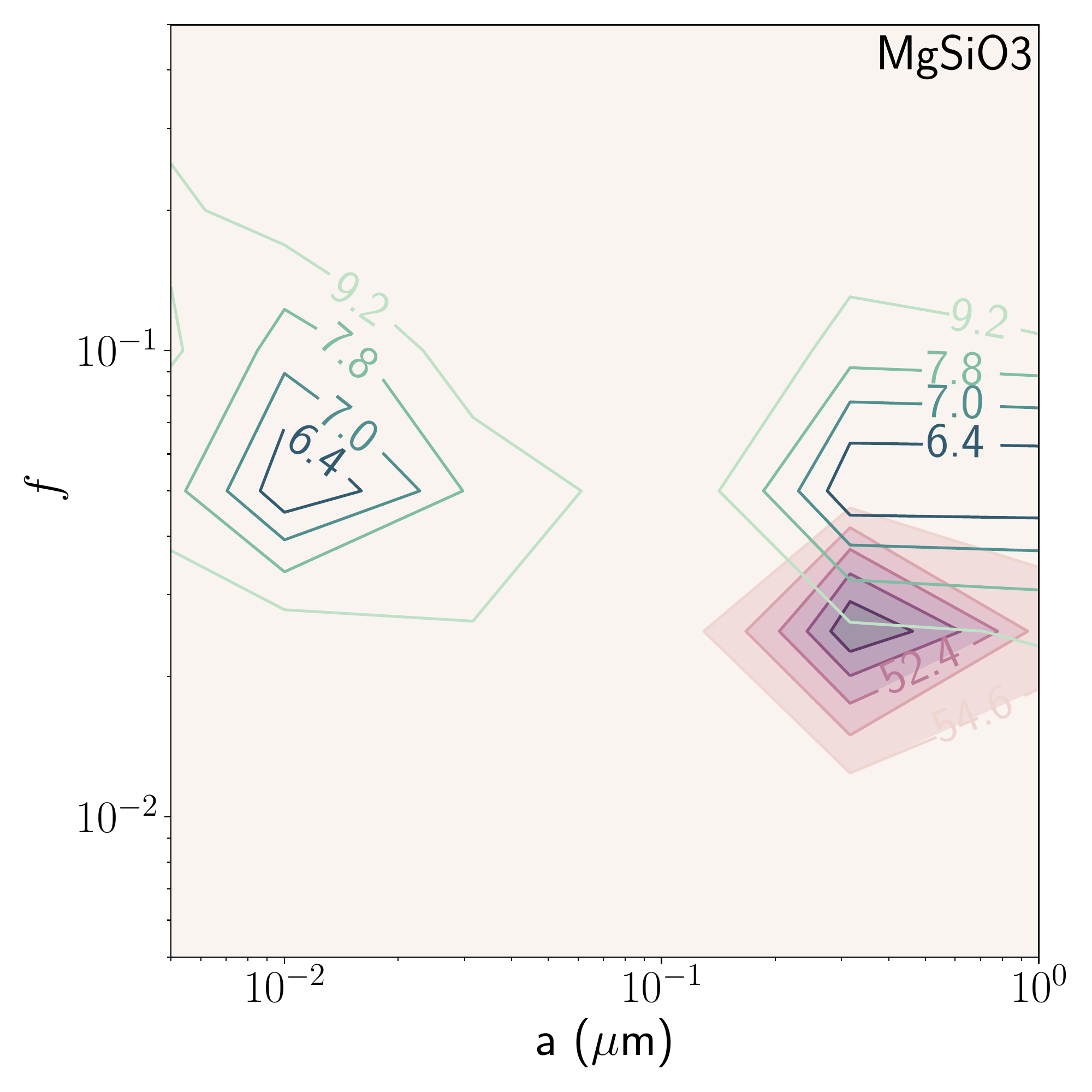}

\hspace{1pt}
\includegraphics[width=0.32\textwidth, clip = True, trim = 10 0 10 0]{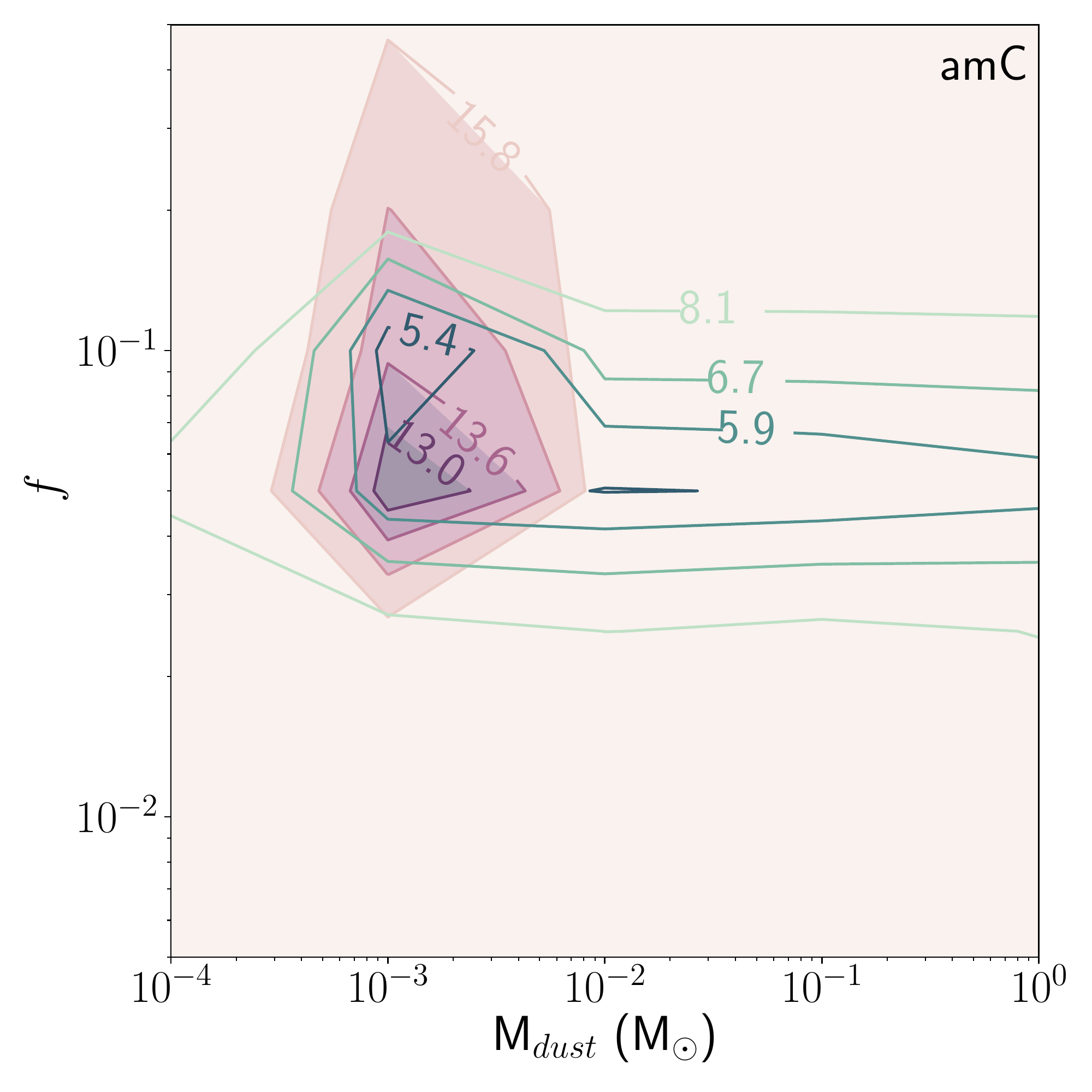}
\includegraphics[width=0.32\textwidth, clip = True, trim = 10 0 10 0]{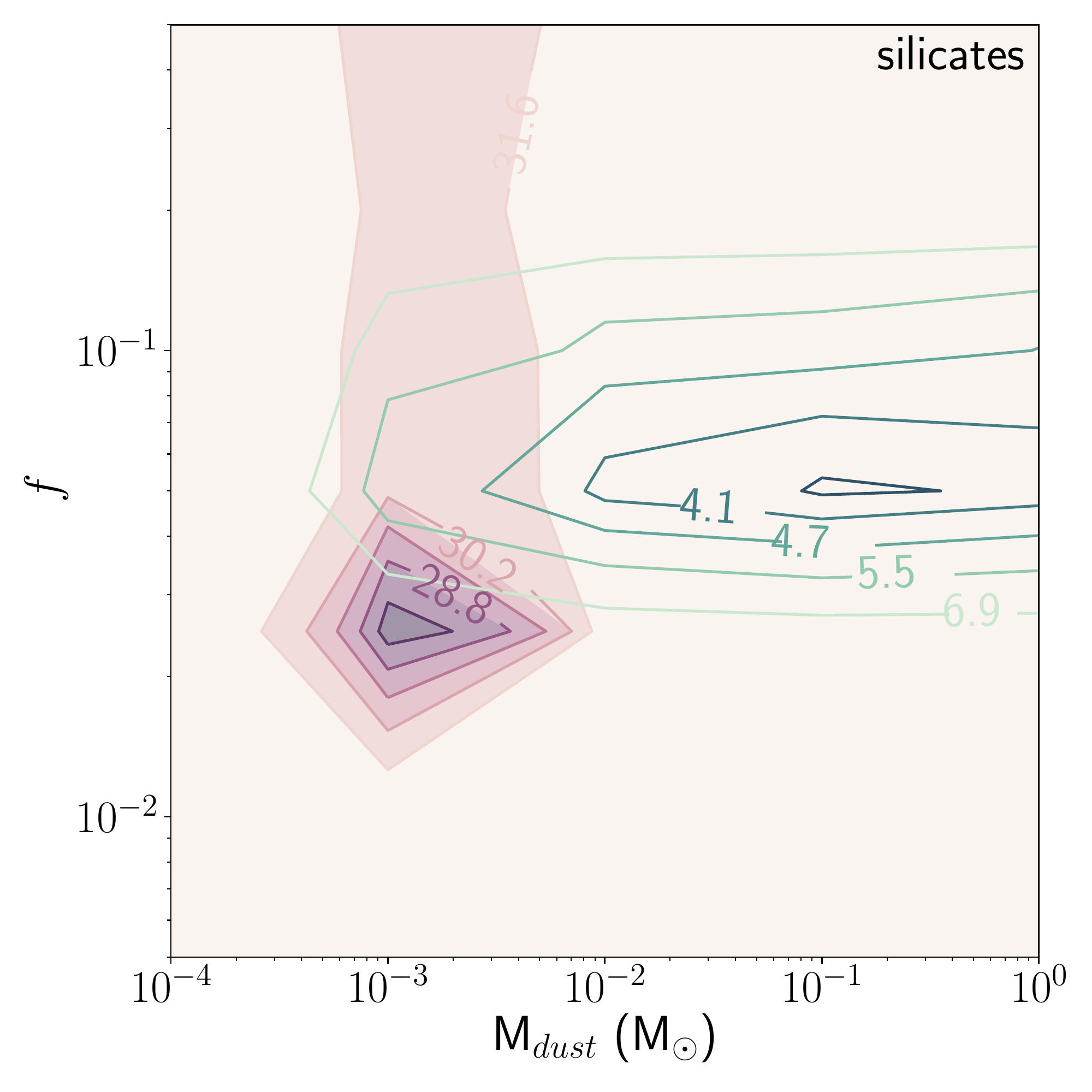}
\includegraphics[width=0.32\textwidth, clip = True, trim = 10 0 10 0]{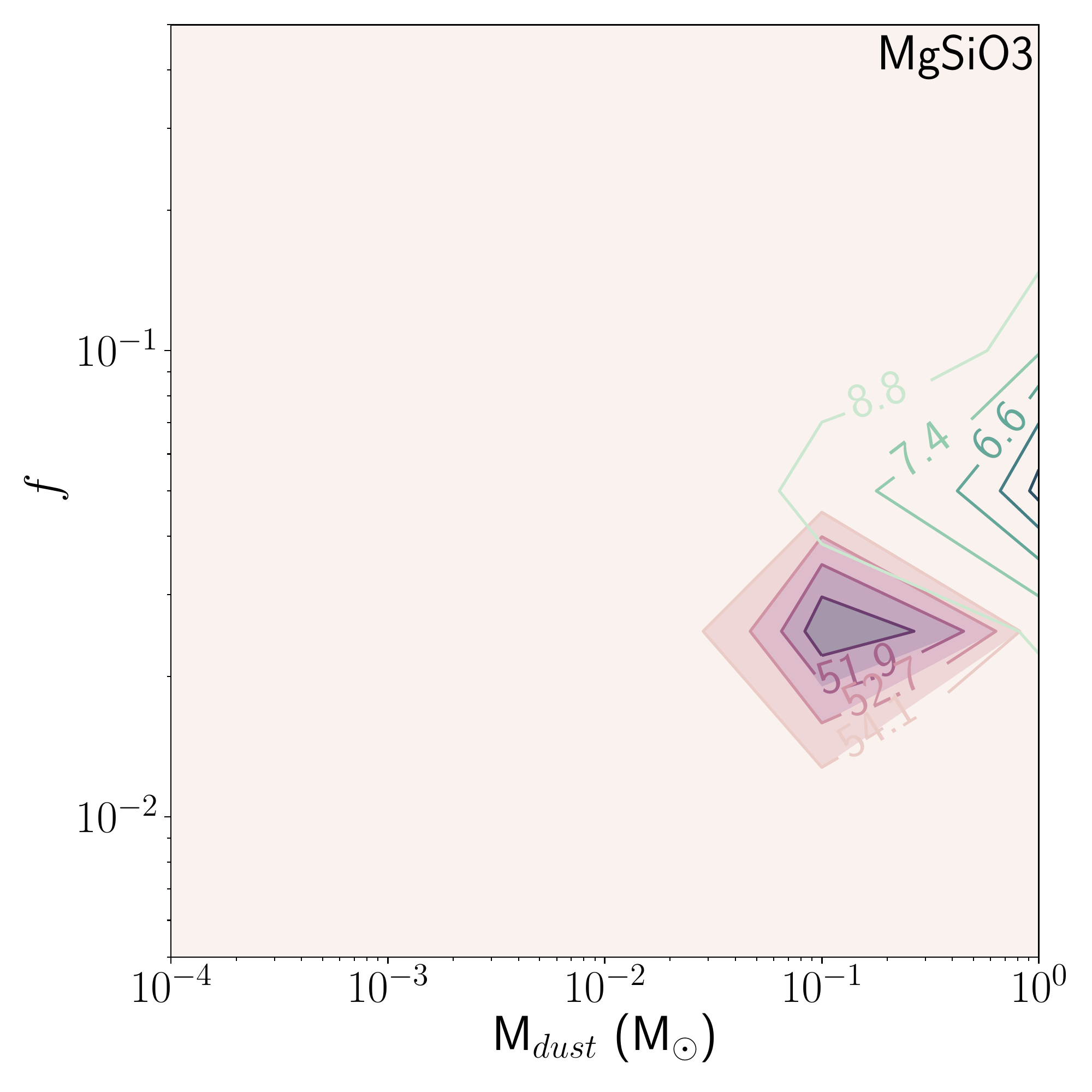}

\caption{2D normalized likelihood contours for the SED models in purple with overplotted 2D normalized likelihood contours for the line profile models in green. Corresponding $\chi^2$ values are given on the contours.}
\label{ff_vs}
\end{figure*}

We initially consider the best-fitting emission line profile models as those models that have $\chi^2<2.5$. We present the variation of dust mass against $\chi^2$ for these models in Figure \ref{Mdust_vs_chi2}.

Whilst both amorphous carbon and silicate grains can achieve good fits to the profile, there are very few MgSiO$_3$ models that are able to achieve good fits to the profile. The few MgSiO$_3$ models that do reproduce the observed profile all share the property of a steep dust clump number density distribution ($n \propto r^{-4}$) and a high dust mass. This geometry means that the majority of photons are emitted in regions with a very high dust density since the emissivity and dust distributions are similar ($n_{dust} \propto r^{-4}$ compared to $n_{[OI]} \propto r^{-4.8}$). Photons therefore get trapped in these highly dense, scattering regions which, despite a very high albedo ($\sim$0.99), results in photons experiencing so many interactions with the dust that they are eventually absorbed. However, whilst there are a handful of potentially viable MgSiO$_3$ emission line models, these parameters give SEDs that are strongly inconsistent with the observed photometry.



Both amorphous carbon grains and silicate grains can reproduce the line profile for a range of dust masses. Across all dust masses, smaller filling factors 0.025, 0.05 and 0.1 are preferred. No models with f=0.005 resulted in good fits since the covering factor to which the line photons are exposed is too small to significantly affect the profile shape. Larger filling factors of 0.2 and 0.5 also failed to produce good fits to the profile. However, based on the previous work of \citet{bevan2016}, we expect that this is likely due to the necessary sparsity of the sampled dust masses in our grid. We note, for example, that the closest parameter set in our amorphous carbon grid to the work of \citet{bevan2016} ($f=0.5$, $R_{clump}=R_{out}/30$, $n_{clump} \propto r^{-2}$ and $a=0.316\mu$m), results in too little absorption for $M_{dust}=0.0001$\,M$_{\odot}$ and far too much absorption for $M_{dust}=0.001$\,M$_{\odot}$.

To mitigate against the limitations imposed by considering only the best-fitting models in a relatively coarsely sampled grid, we have calculated the 2D and 1D marginalized likelihood distributions for the entire grid for both the emission line profile models and SED models for each species (see Figures \ref{corner_amC} - \ref{corner_MgSiO3}). The 2D likelihood contours reveal the correlations between different parameters.  For amorphous carbon grains, we can see that the most favorable volume filling factor (marginalized over the other parameters) is 0.025-0.1. The 2D contour plot in the second panel (first row, second column) of Figure~\ref{corner_amC} also clearly shows a `spike' in the contours that shows, whilst dust masses of $\sim0.001$\,M$_{\odot}$ are generally preferred, the higher dust masses are only accommodated for a particular volume filling factor. This is in agreement with previous assumptions regarding volume filling factors based on Rayleigh-Taylor instabilities \citep{wesson2015}.

However, whilst volume filling factors in this range are broadly favored irrespective of dust mass, the two parameters are strongly linked. In particular, high amorphous carbon dust masses are only able to reproduce the profile when they are restricted to occupy a specific volume filling factor. This is in line with previous work exploring the possible presence of large masses of dust prior to 1000 days in core-collapse supernovae, which have found that the observed line profiles allow large dust masses to be present only if the dust is restricted to a small volume (\citealt{bevan2019}). If the dust is restricted to too small a volume, however, the covering factor of the dust becomes too low to reproduce the blueshifted asymmetry observed in the line profile. In the case of SN~1987A, volume filling factors of $\sim$0.05 are required in order to hide large amounts of dust in the ejecta. This is the single most important parameter in determining goodness-of-fit of the line profile models with large amorphous carbon dust masses $>0.01$\,M$_{\odot}$.

We have adopted the volume filling factor in this paper due to its natural prescription in the models. However, it is the covering factor (i.e. the net surface area of the dust to which the radiation is exposed) that controls the extinction of the dust. This is predominantly determined by the volume filling factor, but is also affected by the clump radius and clump number density distribution. In our models, we have also found that steep clump number density distributions with the majority of the dust residing in the center of the ejecta tend to reproduce the observed [O~{\sc i}] emission line profile better. This is particularly important for the higher dust mass models ($>0.1$\,M$_{\odot}$) since it reduces the covering factor.

Silicate dust models in general require a higher dust mass than amorphous carbon ($\sim0.01 - 0.1$\,M$_{\odot}$) as long as the dust is restricted to a filling factor of $\sim f=0.05$ and a clump number density distribution of $n \propto r^{-2}$. Steeper distributions with the dust more concentrated in the central regions required less dust ($0.001$\,M$_{\odot}$).The likelihood distributions also showed that both amorphous carbon and silicate models were more likely to reproduce the observed [O~{\sc i}] emission with larger grain radii $0.1 - 1.0\mu$m.

Overall, the best-fitting model required 0.01~M$_{\odot}$ of silicate dust and R$_c$=R/15, $f$=0.1, $\rho\propto r^{-2}$ and $a$=0.1$\mu$m. The best-fitting amorphous carbon model required 0.001~M$_{\odot}$ of amorphous carbon dust and R$_c$=R/45, $f$=0.1, $\rho\propto r^{-2}$ and $a$=1.0$\mu$m. 



Whilst there are a number of dust geometries and masses that are able to reproduce the observed  emission line profile accurately, a consistent set of parameters that fits both the SED and the  [O~{\sc i}]~6300,6363~\AA\ doublet at $\sim800$ days is required. We discuss this in the next section.

\section{Searching for models that reproduce both the SED and the emission line profile}
\label{mocassin and damocles}
A viable model of the dust in SN~1987A at $\sim$800 days must reproduce both the observed SED and the emission line profiles. In this section, we search our parameter space for the configurations that come closest to doing this. We first show that the SED models and the line profile models exhibit different dependencies on the parameters. For each species investigated, we calculated the 2D and 1D likelihood distributions based on the $\chi^2$ values by marginalizing over the other parameters. In Figures \ref{corner_amC}, \ref{corner_sil} and \ref{corner_MgSiO3}, we plot these 2D and 1D likelihood distributions for both the line profile models and SED models. This not only illustrates the preferred regions of the parameter space, but also how these regions are different between the line profile models and the SED models. We emphasize that computational constraints limited our ability to obtain a fine sampling of the entire space but we expect the likelihood surface to vary smoothly between our sampled points. We also emphasize that, whilst the plots have been normalized in order to clearly indicate the contours of the likelihood distributions, the regions of highest likelihood do not necessarily correspond to good fits, only to the best fits available in our parameter space; the best-fitting models identified by \citet{wesson2015} and \citet{bevan2016} have dust masses between the values sampled in our grid of models.

The 1D likelihood distributions on the diagonals of Figures \ref{corner_amC}, \ref{corner_sil} and \ref{corner_MgSiO3} give an initial indication of where the SED and emission line profile models are in disagreement. In particular, the MgSiO$_3$ models are discrepant in their required dust masses, whilst also requiring difference volume filling factors and clump radii. Similarly, the silicate models are discrepant in their required dust masses, filling factors and grain sizes. There is much better agreement between the SED and line profile models for amorphous carbon models.

In these figures, one can see that for emission line profiles, it is possible to find a specific small range of filling factors for which the dust mass becomes a lower limit. Analyzing one emission line at one epoch does not allow this degeneracy to be broken, but SED predictions remain sensitive to the mass. In Figure \ref{ff_vs}, we highlight the key 2D contour plots that best illustrate where our models are in agreement and disagreement considering filling factor against grain radius, clump radius and dust mass. We overplot the contours of the line profile model marginalized 2D likelihood distributions on the SED model likelihood contours.

For amorphous carbon grains, there is significant overlap of the most likely regions of parameter space likelihood for the SED and emission line profile models.  A dust mass of $\sim 2 \times 10^{-3}$\,M$_{\odot}$ with a filling factor of $f=0.05$, clump radius of 0.035R$_{out}$ and grain radius $a=0.4\mu$m represents the most likely region, accounting for both models. This is in strong
 agreement with the work of both \citet{wesson2015} and \citet{bevan2016} who found values of 2$\times$10$^{-3}$\,M$_{\odot}$ with a filling factor of $f=0.1$, clump radius of 0.033R$_{out}$, and $\sim 4 \times 10^{-3}$\,M$_{\odot}$ with a filling factor of $f=0.1$, clump radius of 0.02R$_{out}$, respectively.

For silicate grains, there is no agreement between the regions of highest likelihood for grain radius, filling factor and clump radius. Of particular importance, the regions of highest likelihood for dust mass are strongly discrepant. The regions of highest likelihood for the dust mass center around 0.1 for the emission line profile models compared to 0.001 for the SED models. The best compromise between these regions lies at around $3 \times 10^{-3}$\,M$_{\odot}$.

For MgSiO$_3$, the highest likelihood regions of parameter space are mostly discrepant, especially for dust mass. This indicates that there is no set of parameters (within the bounds of our grid) for which viable MgSiO$_3$ dust models can reproduce both the SED and the observed [O~{\sc i}]~6300,6363\AA\ line profile at $\sim$800 days. We also emphasize here that MgSiO$_3$ models are strongly disfavored by the SED fits alone and that, whilst these contour plots indicate regions of highest likelihood, that does not mean that the model is capable of reproducing the observations well.





\section{Discussion}

Considering observations of SN~1987A $\sim$800 days after explosion, we have investigated a large range of parameters encompassing representative values for core-collapse supernova ejecta, with dust masses from 10$^{-4}$ to 1~M$_\odot$, with the aim of determining whether any plausible clump geometry exists which can hide large masses of dust at relatively early times. The different sensitivities of the SED and the emission line profile to the model parameters allow us to clearly constrain the regions of our parameter space where the most plausible models lie. We find that, taking into account both the SED and the emission line profiles, a mass of $\sim$ 10$^{-3}$ M$_\odot$ of amorphous carbon dust is strongly preferred; for higher or lower dust masses, the vast majority of the models investigated across a wide variety of clump configurations result in poor fits, especially to the SEDs.

We further find that for amorphous carbon dust, both the SED and emission line profiles are best reproduced with small filling factors of $f\sim0.05$. Larger dust grains, steeper clump number density distributions, and smaller clumps are all generally preferred. For other dust species, the best-fitting SED and emission line profiles occupy different regions of parameter space, and in particular require dust masses that differ by two orders of magnitude for silicate grains and at least one order of magnitude for MgSiO$_3$.


We find that clump geometries exist that can reproduce the day 775 SED fairly well with 0.01~M$_\odot$ of amorphous carbon dust, a factor of five higher than the best-fitting model presented by \citet{wesson2015}, and 13 times higher than the models presented by \citet{ercolano2007a}. This higher mass remains a factor of 50--100 below the masses estimated using the most recent observations of SN~1987A, 23+ years after its explosion (\citealt{matsuura2011,matsuura2015}). The model is also severely inconsistent with the observed [O~{\sc i}]~6300,6363\AA\ doublet: the parameters that give a good fit to the SED give an extremely poor fit to the emission line profile.

Similarly, dust geometries exist that can hide a large mass of dust from the [O~{\sc i}] line profile, but only for a very specific value of the covering factor, corresponding to a clump volume filling factor of $f\sim0.05$. Excepting this case, the dust mass can always be constrained by the line profile and the modeling presented by \citet{bevan2016} stands, with dust masses of a few 10$^{-3}$\,M$_{\odot}$ providing the best fits to the observations, consistent with the SED modeling of \citet{wesson2015}. Considering both sets of models together, Figure~\ref{ff_vs} shows that only for amorphous carbon dust can the SED and the emission line profiles both be reproduced with the same dust parameters. Carbon dust with a mass of $\sim 2 \times 10^{-3}$\,M$_{\odot}$, a filling factor of $f=0.05$, clump radius of 0.035R$_{out}$ and grain radius $a=0.4\mu$m is thus the best-fitting parameter set, accounting for both models. 

This work therefore further supports the scenario in which dust formation continues for many years after the supernova explosion. The strong upper limit of 3$\times$10$^{-3}$\,M$_\odot$ $\sim$1000 days after the explosion found by \citet{wesson2015} remains applicable, and the vast majority of the $\sim$0.5\,M$_\odot$ of dust now present in the remnant must have formed during the period between then and the Herschel Space Observatory observations of 2011.

We have only considered pure dust compositions in this paper. In reality, models of dust formation predict that a number of different species are formed and coexist within the ejecta (\citealt{sarangi2015,sluder2018}). The computing time that would have been required to extend our analysis to cover mixtures of species was unfortunately prohibitive, but as silicate models in general require a higher dust mass to come close to matching the observations, a model with a carbon-silicate mixture will require a slightly higher dust mass than a pure carbon model. Models in which different dust species have different spatial distributions might provide some scope for fitting the SED with higher dust masses than we have found possible.

\begin{figure*}
    \includegraphics[width=\textwidth]{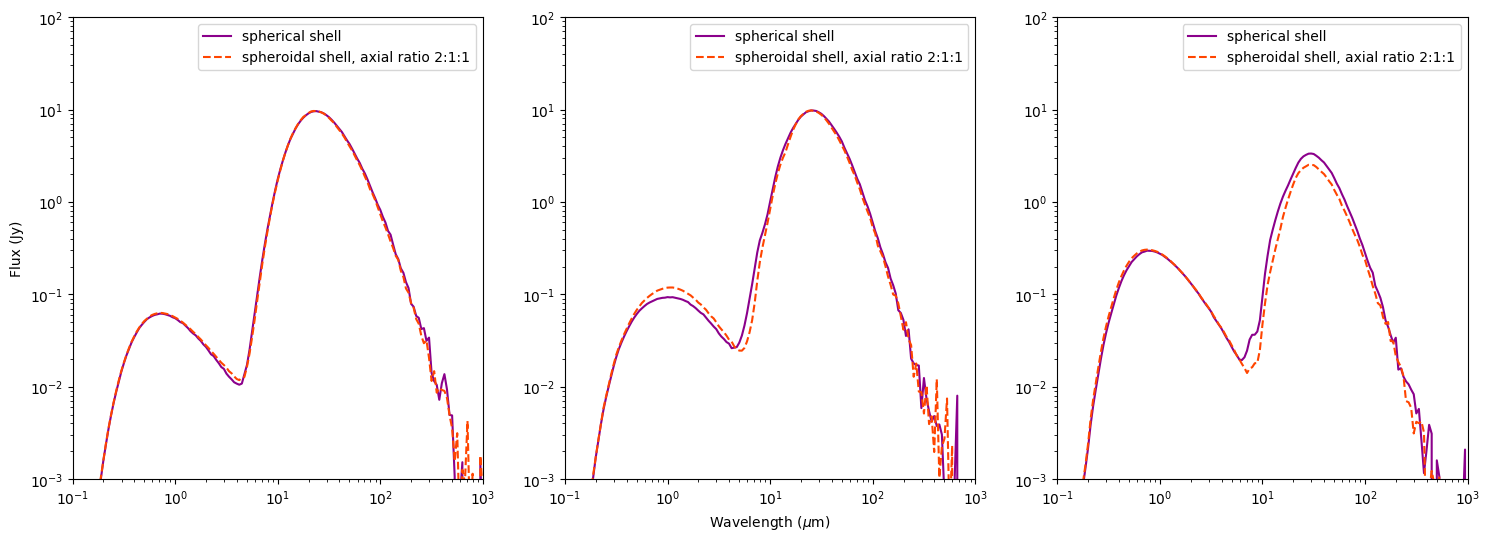}
    \caption{Predicted SEDs for models with all parameters identical except for the dust geometry. The solid line shows the SED for a spherical shell, while the dashed line shows the SED for a spheroidal shell with axial ratio 2:1:1. The models all have M$_{\rm d}$=0.01\,M$_\odot$, $f$=0.1, R$_{\rm c}$=R$_{\rm out}$/30 and $a$=0.1$\mu$m, and are calculated for (left to right) amorphous carbon, astronomical silicates, and MgSiO$_3$.}
    \label{spheroidmodel}
\end{figure*}

Another significant consideration that we have not explored in this paper is ejecta asymmetry. All of our models are spherically symmetric, while in reality, supernova ejecta may not be so simple. In SN~1987A, \textit{Hubble Space Telescope} observations in recent years show that the ejecta is elongated with a dark central region, perhaps corresponding to the hollow center of the expanding shell, or perhaps caused by dust extinction (\citealt{larsson2013}, \citealt{larsson2016}, \citealt{cigan2019}). However, while asymmetric ejecta would have significantly different emission line profiles to symmetric ejecta, the effect on the SED is fairly small. We show this in Figure~\ref{spheroidmodel}, which compares SEDs predicted for spherically symmetric models with those predicted for a spheroidal dust distribution with axial ratios of 2:1:1. The models all have $f$=0.1, R$_{\rm c}$=R$_{\rm out}$/30, $\rho \propto r^{-2}$, M$_{\rm d}$=0.01\,M$_\odot$ and grains with a radius of 0.1$\mu$m. The figure shows that for amorphous carbon dust, the difference between the SEDs is minimal, while for silicate species, the difference is such that changing the dust mass by $\sim$20\% might result in a better fit to the SED. Ejecta asymmetries thus do not offer a way to conceal much greater dust masses than we have found plausible in this study.

Clearly, a natural consideration of emission line modeling is the geometry of the intrinsic emissivity distribution. An inherently asymmetric emissivity distribution could be the cause of an observed asymmetry in the line profile. However, while in this work we have only considered a single epoch of observations, models of the dust mass evolution must also satisfy the observed evolution of the SED and emission line profiles. In SN~1987A, the line profiles are initially symmetrical before gradually becoming asymmetrical over time, indicative of the gradual formation of dust (\citealt{bevan2016}). This evolution has also been seen in several other supernovae where the necessary observational data exists; for example, SN\,1995N from around 700-1400 days after explosion (\citealt{fransson2002}) and SN\,2005ip from 48 to 4075 days after discovery (\citealt{bevan2019}). Because of the time evolution of the blueshifted asymmetry, we infer that it is the result of dust and not of an inherent asymmetry, which would have been seen in early-time spectra prior to 600 days. The possibility that the asymmetry observed in the 806 day [O~{\sc i}] 6300,6363\AA~doublet is induced by concealed, large dust masses relies on a very specific clump configuration. Given that this gradual evolution from symmetric to asymmetric blueshifted line profiles has been observed for several IR and optical lines for several supernovae (e.g. \citealt{stritzinger2012, bevan2016, bevan2020}), and that other supernovae have been shown to require different filling factors to conceal large dust masses from the line profile, it increasingly seems implausible that all of these objects coincidentally exhibit the required clump configuration.


We therefore conclude that, if a geometry exists that does allow large dust masses to be present but to have an unobservable effect on both the SED and the emission lines, it lies outside the range of parameters we have considered. Unless such a geometry can be identified, we infer that the majority of dust in SN\,1987A formed many years after the explosion. Given the similar evolution with time of emission line profiles seen in other remnants, it seems likely that this finding applies generally, and theoretical models of dust formation will therefore need to reproduce this behavior.

\section{Acknowledgements}

RW and AB are supported by European Research Council Grant SNDUST 694520. We thank Florian Kirchschlager, Ilse de Looze, and Mike Barlow for providing detailed comments on the draft. We thank the anonymous referee for their report that helped us to significantly improve the draft.


\bibliographystyle{aasjournal}
\bibliography{references} 

\end{document}